\documentclass[11pt,a4paper]{article}
\usepackage{jcappub}

\pdfoutput=1

\usepackage{graphicx}
\usepackage{epsfig}
\usepackage{rotate}
\usepackage{amsmath}
\usepackage{amssymb}
\usepackage{amsfonts}
\usepackage{enumerate}
\usepackage{afterpage}
\usepackage{placeins}
\usepackage{caption}
\usepackage{xcolor}
\usepackage[T1]{fontenc}
\usepackage{booktabs}
\usepackage{braket}
\usepackage{subfig}
\usepackage{todonotes}
\usepackage[utf8]{inputenc}
\hypersetup{
 citecolor=blue,filecolor=blue,linkcolor=blue,urlcolor=blue}
\usepackage{array}
\newcolumntype{P}[1]{>{\centering\arraybackslash}p{#1}}
\newcolumntype{M}[1]{>{\centering\arraybackslash}m{#1}}

\allowdisplaybreaks[1]

\def\ra{\rightarrow}
\newcommand{\by}{{\mathbf y}}
\newcommand{\bx}{{\mathbf x}}

\newcommand{\bn}{{\mathbf n}}

\newcommand{\bk}{{\mathbf k}}

\newcommand{\br}{{\mathbf r}}

\newcommand{\pa}{{\parallel}}

\newcommand{\PP}{{\cal P}}

\newcommand{\HH}{{\cal H}}
\newcommand{\cd}{\cdot}

\newcommand{\de}{\delta}
\newcommand{\De}{\Delta}

\newcommand{\ga}{\gamma}

\newcommand{\La}{\Lambda}
\newcommand{\la}{\lambda}
\newcommand{\Om}{\Omega}

\newcommand{\be}{\begin{equation}}
\newcommand{\ee}{\end{equation}}
\newcommand{\bea}{\begin{eqnarray}}
\newcommand{\eea}{\end{eqnarray}}
\newcommand{\bean}{\begin{eqnarray*}}
\newcommand{\eean}{\end{eqnarray*}}
\newcommand{\beal}{\begin{align}}
\newcommand{\enal}{\end{align}}
\newcommand{\pd}{\partial}
\newcommand{\dd}{\text{d}}

\newcommand{\kcos}{\nu}
\newcommand{\nuP}{\nu}

\usepackage{ntheorem}

\newtheorem*{theorem-non}{Theorem}

\newcommand{\mathleft}{\@fleqntrue\@mathmargin0pt}
\newcommand{\mathcenter}{\@fleqnfalse}

\makeatletter
\newcommand*{\centerfloat}{%
  \parindent \z@
  \leftskip \z@ \@plus 1fil \@minus \textwidth
  \rightskip\leftskip
  \parfillskip \z@skip}
\makeatother

\definecolor{revisioncolor}{RGB}{22, 158, 230}

\newcommand{\code}[1]{{\sc #1}} % for names of codes and programming languages
\newcommand{\source}[1]{\texttt{#1}} % for actual names of variables/functions in source code

\newcommand{\tj}[6]{ \begin{pmatrix}
   #1 & #2 & #3 \\
   #4 & #5 & #6 
\end{pmatrix}}

\begin{document}
\title{{\sc coffe}: a code for the full-sky relativistic galaxy correlation function}
\author{Vittorio Tansella,  Goran Jelic-Cizmek, Camille Bonvin and Ruth Durrer}
\affiliation{D\'epartement de Physique Th\'eorique and Center for Astroparticle Physics, Universit\'e de Gen\`eve, 24 Quai Ansermet, CH--1211 Gen\`eve 4, Switzerland}
\emailAdd{vittorio.tansella@unige.ch}
\emailAdd{goran.jelic-cizmek@unige.ch}
\emailAdd{camille.bonvin@unige.ch}
\emailAdd{ruth.durrer@unige.ch}
\vspace{1 em}
\date{\today}

\abstract{We present a public version of the code {\sc{coffe}} (COrrelation Function Full-sky Estimator) available at \url{https://github.com/JCGoran/coffe}. The code computes the {galaxy two-point} correlation function and its multipoles in linear perturbation theory, including all relativistic and wide angle corrections. {\sc coffe} also calculates the covariance matrix for two physically relevant estimators of the correlation function multipoles.  We illustrate the usefulness of our code by a simple but relevant example: a forecast of the detectability of the lensing signal in the multipoles of the two-point function. In particular, we show that lensing should be detectable in the multipoles of the two-point function, with a signal-to-noise larger than 10, in future surveys like Euclid or the SKA.}

\maketitle

\section{Introduction}
The two-point function of galaxies contains valuable information about cosmology and the large-scale structure (LSS) of the universe. Measurements of the two-point correlation function (2pF) have been performed by different collaborations over the past years~\cite{Jones:2009yz,2013AAS...22110602D,2015ApJS..219...12A} and upcoming redshift surveys will probe the LSS of the universe at deeper redshift and for larger volumes~\cite{2017ApJS..233...25A,2011arXiv1110.3193L,Maartens:2015mra}. It is often argued that cosmology has become a precision science thanks to the very accurate measurement of the Cosmic Microwave Background (CMB) temperature  fluctuations and polarization power spectra~\cite{Adam:2015rua,Ade:2015xua} and it is now time for the observation of galaxy distribution to contribute to this name. It is clear that to  correctly interpret and to profit maximally from the data that will soon be available we need robust theoretical predictions of the signal. Not only the signal has to be understood from a theoretical point of view but it is necessary to find accurate and fast methods to compute it. For the CMB, we have at our disposal fast linear Boltzmann codes such as {\sc camb}~\cite{Lewis:1999bs} and {\sc class}~\cite{Blas:2011rf}. Recently, these codes have been extended to compute also the angular power spectrum of galaxy number counts, $C_\ell$,~\cite{2011PhRvD..84d3516C} and~\cite{DiDio:2013bqa}. However, redshift surveys traditionally measure the 2pF and its multipoles (monopole, quadrupole and hexadecapole), rather than the $C_\ell$'s. In this work we present a public version of the code {\sc{coffe}} (COrrelation Function Full-sky Estimator) which computes the galaxy 2pF including all the relativistic projection effects and does not rely on the flat-sky approximation. It is known that the galaxy 2pF is not simply given by density fluctuations and redshift-space distortions (RSD) but it acquires several additional terms from lensing, ordinary and integrated Sachs Wolfe effects, gravitational redshift, Doppler terms, and Shapiro time delay. These contributions arise in the expression for the observed galaxy number counts $\De(\bn,z)$ at redshift $z$ and direction $\bn$ in the sky~\cite{2009PhRvD..80h3514Y,Bonvin:2011bg,Challinor:2011bk}  and of course they contribute also to the correlation function defined as
\be
\xi(\cos\theta,z_1,z_2)= \langle \De(z_1,\bn_1) \De(z_2,\bn_2) \rangle\, ,
\label{2pfdef}
\ee
where $\cos\theta = \bn_1\cdot\bn_2$. The brackets in eq.~(\ref{2pfdef}) are intended, from a theoretical point of view, as an ensemble average but if ergodicity holds (as it does for the case of a statistically homogeneous Gaussian random field), in observations, they can be replaced by a spatial average. The expression that is most commonly used in the literature for the 2pF is the Fourier transform of the Kaiser formula for the galaxy power spectrum~\cite{Kaiser1987}
\be\label{e:poldk}
P(\bar z,k, \nuP)_\text{\,Kaiser} = D^2_1(\bar z)\left[b^2+\frac{2bf}{3}+\frac{f^2}{5}+\left(\frac{4bf}{3}+\frac{4f^2}{7} \right)\PP_2(\kcos)
+\frac{8f^2}{35} \PP_4(\kcos)\right]P(k)\, .
\ee
Here $\bar z$ is the mean redshift of the survey, $P(k)$ is the matter density power spectrum today,  $D_1(\bar z)$ is the growth factor normalised to $1$ today, $\nuP$ is the cosine of the angle between $\bk$ and the line-of-sight direction (assumed fixed in the flat-sky limit), $\nuP=\bn\cdot\bk$, and the $\PP_\ell$ are the Legendre polynomials of degree $\ell$. We have also defined $b(\bar z)$ as the galaxy bias which relates the galaxy density fluctuations to the matter perturbation in synchronous-comoving gauge: $\De^\text{den} = b \cdot \de_c$. Furthermore
\be\label{e:growth}
f(\bar z) = -\frac{D_1'}{D_1}(1+\bar z)=\frac{d\ln D_1}{d\ln(a)} \,,
\ee
is the growth rate, where the prime denotes the derivative with respect to the redshift $\bar z$. eq.~(\ref{e:poldk}) relies on the flat-sky approximation and does not include all the projection effects we mentioned before. A simple way to write  the 2pF in full generality is to use the well know expression
\be\label{e:xiCl}
\xi(\theta,z_1,z_2)= \frac{1}{4\pi} \sum_\ell (2\ell+1)C_\ell(z_1,z_2)\PP_\ell(\cos\theta) \,,
\ee
where $C_\ell(z_1,z_2)$ is the number counts angular-redshift power spectrum introduced in~\cite{Bonvin:2011bg,Challinor:2011bk}. Even though fast and reliable codes such as {\sc camb} and {\sc class} have been generalised to calculate the number count angular power spectrum~\cite{DiDio:2013bqa,DiDio:2013sea}, the use of eq.~(\ref{e:xiCl}) to compute the 2pF is not advisable. As explained in \cite{Campagne:2017wec,Tansella:2017rpi} this approach has two relevant drawbacks:
\begin{itemize}
\item \emph{Window function}: eq.~(\ref{e:xiCl}) is essentially an inverse Fourier-Bessel transform. The sum over $\ell$ runs to infinity and we are forced to cut it at some $\ell_\text{max}$. This is equivalent to introducing a top-hat window function $W_\ell$ in $\ell$-space which enforces $C_\ell=0$ for $\ell>\ell_\text{max}$. The inverse transform is then a convolution of $\xi$ with the inverse transform of $W_\ell$ (given usually in terms of spherical Bessel functions $j_\ell$): this introduces spurious oscillations in the result. A possible workaround is to introduce in the sum a decaying window function which ensures $(2\ell+1) C_\ell \simeq 0$ for $\ell\,\gtrsim\, \ell_\text{max}$ but the result will then depend on the smoothing scale chosen.
\item \emph{Run time}: Typical values for $\ell_\text{max}$ in order to reproduce the correct behaviour of $\xi$ are $\ell_\text{max} > 3000$. This means that every point of the 2pF requires the computation of several thousands spectra $C_\ell$: {\sc class} is very fast but this quickly becomes unfeasible, especially when terms which require line-of-sight integrations are sought (i.e. lensing).
\end{itemize}
As argued in Ref.~\cite{Tansella:2017rpi}, this problems become especially relevant when we want to exploit the very high redshift resolution of spectroscopic surveys in a redshift bin $\De z$. For correlating only a small number of rather wide photometric redshift bins, the $C_\ell(z_1,z_2)$  probably remain the method of choice.

The code {\sc coffe}  performs a direct calculation of the 2pF which does not need the angular power spectra $C_\ell(z_1,z_2)$.  For the standard and one of the Doppler terms  this has already been done in~\cite{Campagne:2017wec}, with an implementation in the public code $\source{AngPow}$ \cite{Campagne:2017xps}. Here we extend this work to include lensing and all other relativistic effects in full sky. {In particular, our code computes: 1)~the 2pF as a function of redshift, separation and orientation, 2)~the multipoles of the 2pF, which is the output directly delivered by redshift surveys, and 3)~the covariance matrix, necessary e.g. to assess the detectability of lensing and relativistic effects and their information content.} In the next section we summarise the theoretical results which allow for a direct calculation and we deal with the problem of Infra-Red (IR) divergence which afflicts some terms in the correlation function. In section \ref{application} we illustrate the usefulness of our code through one simple example. In section~\ref{struct} we present the code and in section~\ref{next} we conclude and discuss future implementation to expand its functionalities.

%%%%%%%%%%%%%%%%%%%%%%%%%%%%%%%%%%The full-sky relativistic correlation function%%%%%%%%%%%%%%%%%%%%%%%%%%%%%%%%%%%%%%%%%%%%
\section{The relativistic full-sky correlation function}\label{theo}
\subsection{The formalism}
In this section we summarize the results obtained in~\cite{Tansella:2017rpi}. Let us start with the set up for computing the {two-point} correlation function (2pF). The 2pF is usually not regarded as a function of two redshifts and one angle as in eq.~(\ref{2pfdef})  but as a function of the separation between the two points  $r$, the mean redshift $\bar z$ and the cosine $\mu$ of the angle between  the separation vector $\br$ and a line-of-sight (LOS) between the two directions of observation, determined by convention. There is not a unique way to define this angle. In the flat-sky limit common definitions coincide, but in full sky they lead to differences in the multipoles, which are potentially of the same order of magnitude as the relativistic effects. It is therefore crucial to clearly specify the chosen angle. It is common practice to split  $\br$ into its parallel component $r_\parallel$ (i.e. parallel to the LOS) and transverse component $r_\perp$ (i.e. perpendicular to the LOS) so that $r=\sqrt{r_\parallel^2+r_\perp^2}$. In the full-sky regime, where we take into account that the two points do not share the same LOS, we chose to define the parallel separation as the difference between the comoving distance of the two points
\be
r_\parallel = \chi_2 -\chi_1 \,,
\ee
where $\chi_i=\chi(z_i)$. We also define $\mu$ in the usual way as $\mu = r_\parallel/r$, which reduces to the standard definition in the flat-sky limit. Assuming vanishing spatial curvature $\Omega_K=0$ (as this first release of {\sc coffe} does) the separation between the two points is given by
\be
r(\theta,z_1,z_2) = \sqrt{\chi_1^2+\chi_2^2 -2 \chi_1 \chi_2 \cos\theta} \,,
\label{rexpr}
\ee
and $\cos\theta$ can be related to $r$, $\mu$ and $\bar z$ by
\be
\cos\theta  = \frac{2\bar\chi^2-r^2+\frac{1}{2}\mu^2r^2}{2\bar\chi^2-\frac{1}{2}\mu^2r^2} \,,
\label{e:ctheta}
\ee
where we have introduced\footnote{Note that $\bar\chi$ and $\chi(\bar z)$ are not exactly the same but in what follows we neglect this difference which is of order $(\De z)^2/\HH(\bar z)$.} $\bar\chi=(\chi_1+\chi_2)/2\simeq \chi(\bar z)$. We point out that when writing the correlation function $\xi(r,\mu,\bar z)$ (considering physical distances) a cosmology must be assumed to convert the observed redshifts to $\chi_1$ and $\chi_2$, while $\xi(\theta,z_1,z_2)$ can be directly measured in observations. However the former approach allows us to compute the multipoles of the correlation function which are often useful to break the degeneracy between cosmological parameters~\cite{Chuang:2012qt,Chuang:2012ad,Cuesta:2015mqa,Hinton:2016atz,Wang:2016wjr,Chuang:2016uuz,Zhao:2018jxv}. One then has to be careful when estimating cosmological parameters, taking into account that the data $\xi(r,\mu,\bar z)$ itself depends on them. {This is usually done by introducing rescaling parameters in the correlation function, which are fitted at the same time as cosmological parameters, see e.g.~\cite{2013MNRAS.431.2834X}.}

Having clarified the setup, we can now turn to the computation of the 2pF. Including all the relativistic corrections, the galaxy number counts can be written as~\cite{Yoo:2009au,Bonvin:2011bg,Challinor:2011bk}
\be
\De(\bn,z)=\Delta^{\rm den}+\Delta^{\rm rsd}+\Delta^{\rm len}+\Delta^{\rm d1}+\Delta^{\rm d2}+\Delta^{\rm g1}+\Delta^{\rm g2}
+\Delta^{\rm g3}+\Delta^{\rm g4}+\Delta^{\rm g5}\,.
 \label{e:Degz}
\ee
We can identify the physical meaning of each term in the following way: the standard terms, i.e. the density fluctuations and the RSD term, denoted respectively by $\Delta^{\rm den}$ and $\Delta^{\rm rsd}$, are usually taken into account in galaxy clustering analyses. $\Delta^{\rm len}$ represents the lensing term. $\Delta^{\rm d1}$ is the Doppler contribution, $\Delta^{\rm d2}$ is a velocity term which comes from transforming the longitudinal gauge density into the comoving density. $\Delta^{\rm g1}, \Delta^{\rm g2}$ and $\Delta^{\rm g3}$ are relativistic effects, given by the gravitational potentials at the source. As such they are sometimes called 'Sachs-Wolfe' terms. $\Delta^{\rm g4}$ denotes the so-called Shapiro time-delay contribution and $\Delta^{\rm g5}$ is the integrated Sachs-Wolfe term. Redshift-space expressions for the contributions in Eq~(\ref{e:Degz}) can be found in~\cite{Tansella:2017rpi}, here we will only use the Fourier-Bessel transform of these terms given by
\bea
\De_\ell^\text{den}&=&  b(z) S_D j_\ell( k \chi ) \,,\label{e:den}\\
\De_\ell^\text{rsd}&=&  \frac{k}{\HH} S_V j_\ell''(k \chi)\,, \label{e:rsd}\\
\De_\ell^\text{len}&=& \left(\frac{2 - 5 s}{2}\right)\frac{\ell(\ell+1)}{\chi} \int_0^\chi \dd \la  \frac{\chi-\la}{\la} (S_\phi+S_\psi) j_\ell( k \la) \,,\label{e:len}\\
\De_\ell^\text{d1}&=&  \left(\frac{\dot \HH}{\HH^2}+\frac{2-5s}{\chi \HH}+5 s - f_\text{evo} \right) S_V j_\ell'(k \chi) \,,\label{e:d1}\\
\De_\ell^\text{d2}&=& -(3-f_{\rm evo})\frac{\HH}{k} S_V j_\ell(k \chi)   \,,\label{e:d2}\\
\De_\ell^\text{g1}&=&  \left(\!\!1+\frac{\dot \HH}{\HH^2}+\frac{2-5s}{\chi \HH}+5 s - f_\text{evo} \!\right)\! S_\psi j_\ell(k \chi)  \,,\label{e:g1}\\
\De_\ell^\text{g2}&=& (-2+5s)S_\phi j_\ell(k \chi)  \,,\label{e:g2} \\
\De_\ell^\text{g3}&=& \frac{1}{\HH} \dot S_\phi j_\ell(k \chi)  \,,  \label{e:g3}\\
\De_\ell^\text{g4}&=&  \frac{ 2-5s}{\chi}  \int_0^\chi \dd \la(S_\phi +S_\psi) j_\ell(k \la)     \,, \label{e:g4}\\
\De_\ell^\text{g5}&=&  \left(\frac{\dot \HH}{\HH^2}+\frac{2-5s}{\chi \HH}+5 s - f_\text{evo} \right) \int_0^\chi \dd \la   (\dot S_\phi +\dot S_\psi) j_\ell(k \la) \,.\label{e:g5}
\eea
We define the matter transfer function $S_D$, which relates the primordial power spectrum $\PP_\zeta(k) = A_s (k/k_*)^{n_s-1}$ to the matter power spectrum at redshifts $z_1$ and $z_2$, via
\be
\mathcal{P}_\zeta (k) S_D(k,z_1) S_D(k,z_2) = \frac{k^3}{2 \pi^2} D_1(z_1) D_1(z_2) P(k)|_{z=0} = \frac{k^3}{2 \pi^2} P(k,z_1,z_2)\,.
\ee
In standard $\La$CDM, the velocity and potentials transfer functions are related to $S_D$ through 
\be
S_V = -(\HH f)/k \,S_D\,,
\ee
\be
S_\Phi = S_\Psi = -\frac{ 3 \Omega_m}{2a} \left( \frac{\HH_0}{k}\right)^2 S_D \,,
\ee
\be
S_{\dot \Phi} =  S_{\dot \Psi} =  -\frac{ 3 \Omega_m}{2a} \left( \frac{\HH_0}{k}\right)^2 ( \dot S_D - \HH S_D) \,.
\ee
Furthermore $s$ denotes the magnification bias and $f_\text{evo}$ is the evolution bias.

\begin{figure*}[ht]
\centerfloat
\includegraphics[scale=0.6]{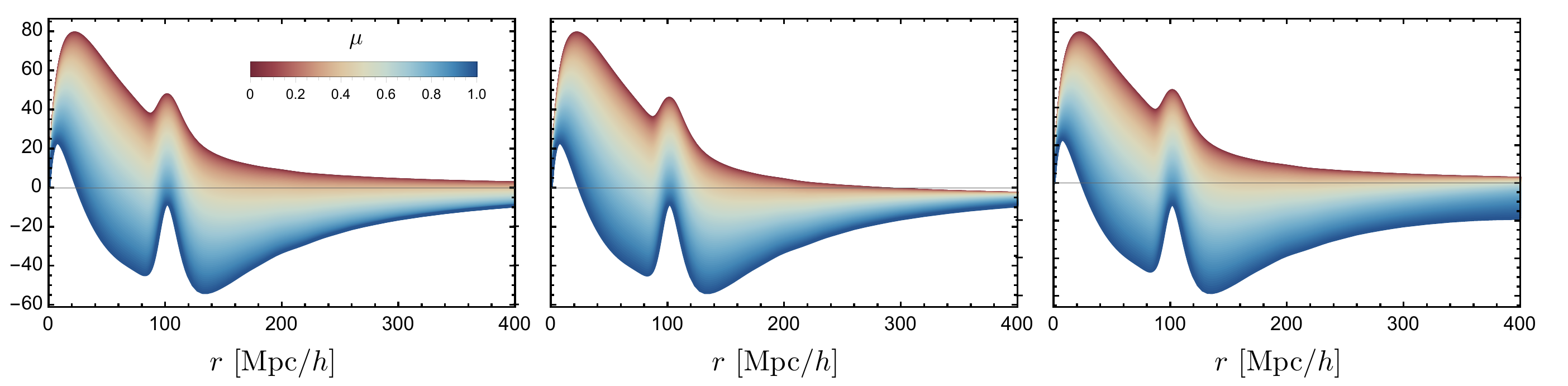}
\caption{\label{f:dem} A possible output of the {\sc coffe} code. The correlation function $r^2 \xi(r,\mu)$ is plotted at $\bar z=0.1$ for different values of $\mu$ (color coded): the \emph{left} panel is the flat-sky result of eq.~(\ref{fsxi}), the \emph{middle} panel the full-sky result considering only density and redshift-space distortion, while the \emph{right} panel also considers the "d1" doppler term (the most relevant relativistic contribution at low redshift). Note that the full sky result for the standard terms is negative at large distances while the flat sky result can be positive depending on the orientation $\mu$. Including the relativistic terms (in this case only d1 contributes visibly), the correlation function becomes again positive for almost transverse orientations.  For large $\mu$ and large $r$ it is significantly more negative than the standard flat sky result.} 
\end{figure*}

The basic idea (envisioned in~\cite{Campagne:2017wec} and exploited in~\cite{Tansella:2017rpi}) upon which the direct calculation is performed is based on eq.~(\ref{e:xiCl}) and the explicit expression for the contributions of the correlation of the terms $A$ and $B$ in the angular power spectrum
 \be\label{e:CABl}
 C_\ell^{AB} (z_1,z_2) \,=\, 4 \pi\hspace{-1.2mm} \int\! \frac{ dk }{k} \mathcal{P}_{\mathcal{R}}(k) \De_\ell^A(k,z_1)\De_\ell^B(k,z_2)\,,
 \ee 
where we set schematically $\{A,B\} = \{\text{den}, \text{rsd},...,\text{g5}\}$. Combining Eqs. (\ref{e:xiCl}) and (\ref{e:CABl}) we can exchange the sum over $\ell$ and the integral over the wavenumber: we then need to evaluate sums of the form
\be
 \sum\limits_\ell (2\ell+1) \De_\ell^A(k,z_1)\De_\ell^B(k,z_2) \PP_\ell(\cos \theta) \,.
 \label{resum}
\ee
The $\De^A_\ell$ depend on $\ell$ only via a spherical Bessel function or derivatives of it and hence we can perform the infinite sum over $\ell$ analytically, leading to simple functions of $\theta$, $z_1$, $z_2$ multiplying $j_L(kr)$ with $L\in\{0,1,2,3,4\}$. We refer the interested reader to section 2.2 of~\cite{Tansella:2017rpi} for details of this calculation. Here we report the results with a somewhat different notation to present the expressions in a way which is closer to what is implemented in {\sc coffe}. It is useful to split the discussion between non-integrated terms $\{ \text{den}, \text{rsd},\text{d1}, \text{d2}, \text{g1}, \text{g2}, \text{g3}\}$ and integrated terms $\{\text{len}, \text{g4} ,\text{g5}\}$, which require a LOS integration.\\

For the non-integrated terms we define\footnote{Note that since we have assumed a cosmological model we write $\xi(\theta,\chi_1,\chi_2)=\xi(\theta,z_1,z_2)=\xi(r,\mu,\bar z)=\xi(r,\mu,\bar \chi)$ with no distinction. In fact we can use $\chi_i=\chi(z_i)$, $\mu = (\chi_2-\chi_1)/r $ and Eqs.~(\ref{rexpr}),(\ref{e:ctheta}) to switch between the different variables in which we express the correlation function. We neglect the difference between $\bar \chi$ and $\chi(\bar z)$. }
\be 
\xi^{AB} (\theta,\chi_1,\chi_2) = D_1(\chi_1) D_1(\chi_2)\sum\limits_{\ell, n} \bigg(X_\ell^n \big|_A+X_\ell^{n}\big|_{AB}+X_\ell^{n}\big|_{BA}+X_\ell^n\big|_B \bigg) I^n_\ell(r)\, ,
\label{xiAB}
\ee
where the $X_\ell^n\big|_{AB} = X_\ell^n (\theta,\chi_1,\chi_2)\big|_{AB}$, $\{A,B\} = \{\text{den}, \text{rsd},\text{d1},\text{d2},\text{g1},\text{g2},\text{g3}\}$ are listed in Appendix~\ref{appcoeff}. Note that a single tag means autocorrelation: $X_\ell^n \big|_A \equiv X_\ell^n \big|_{AA}$. The sum is intended over all the values of $\ell, n \in \{0,1,2,3,4\}$ for which the coefficients (given in the appendix) are non-zero. 
Note that the symmetry of the 2pF implies
\be
X_\ell^n \big|_{AB}(\theta,\chi_1,\chi_2 )= X_\ell^{n}\big|_{BA}(\theta,\chi_2,\chi_1)\,,
\ee
and we have defined 
\be
I_\ell^n(r) = \int \frac{dk\, k^2}{2 \pi^2} \,P(k)\,\frac{ j_\ell( k r )}{(k r)^n} \,.
\ee
The use of this notation is justified in two ways: firstly it is now clear that the integrals $I^n_\ell(r)$ need to be computed only once for every separation, independently of the orientation (i.e. $\mu$). This fact was somewhat hidden in the notation of~\cite{Tansella:2017rpi} and we make it explicit here. Secondly we have isolated the integrals $I^n_\ell (r)$: a fast and accurate computation of these integrals is crucial for the precision of the 2pF. We have implemented the {\sc 2-fast}~\cite{Gebhardt:2017chz} algorithm in {\sc C} and included it in our code\footnote{The original, publicly available, {\sc 2-fast} code (\url{https://github.com/hsgg/twoFAST}) is implemented in the high-level language \textbf{julia}.}.  In eq.~(\ref{xiAB}),  $\xi^{AB}$ means $\xi^{AB} = \langle ( A+B)(A+B) \rangle$ and in general we define 
\be 
\xi^{ABCD...} = \langle ( A+B+C+D+...)(A+B+C+D+...) \rangle\, ,
\label{notxi}
\ee
where in this case the sum in eq.~(\ref{xiAB}) is done over all possible combinations. \\

For the integrated terms we define
\be 
\xi^{AB} (\theta,\chi_1,\chi_2) =  \bigg(Z \big|_A+Z\big|_{AB}+Z\big|_{BA}+Z\big|_B \bigg)\, ,
\label{IxiAB}
\ee
where $Z = Z (\theta,\chi_1,\chi_2)$, $\{A,B\} = \{\text{den},\text{rsd},...,\text{len}, \text{g4},\text{g5}\}$ and a single tag means autocorrelation. We again have 
\be
Z \big|_{AB}(\theta,\chi_1,\chi_2 )= Z\big|_{BA}(\theta,\chi_2,\chi_1) \,,
\ee
and the full list is given in Appendix~\ref{appcoeff}. Examples of the correlation function are shown in fig.~\ref{f:dem}.

For completeness we also give the definition of the multipoles of the correlation function:
\be
\xi_\ell(z,r) \equiv \frac{2\ell+1}{2} \int\limits_{-1}^1 d\mu \, \xi(z,r,\mu) \PP_\ell(\mu) \,,
\label{multdef}
\ee
and we remind the reader that in the flat-sky approximation, redshift-space distortions are included by Fourier transforming eq.~(\ref{e:poldk}), which yields 
\be
\xi(\bar z, r, \mu)_\text{flat-sky} = D^2_1(\bar z) \bigg[c_0(\bar z) I^0_0(r) - c_2(\bar z) I^0_2(r) \PP_2(\mu) +c_4(\bar z) I^0_4(r) \PP_4(\mu)\bigg] \,,
\label{fsxi}
\ee 
with
\begin{align}
c_0 &=  b^2+\frac{2}{3}bf +\frac{f^2}{5} \,, \label{e:c0}\\
c_2 &=  \frac{4}{3} bf +\frac{4}{7} f^2 \,, \\
c_4 &= \frac{8}{35} f^2\,. \label{e:c4}
\end{align}

\subsection{IR divergence}\label{IRdiv}

\begin{figure*}[ht]
\centering
\includegraphics[scale=0.56]{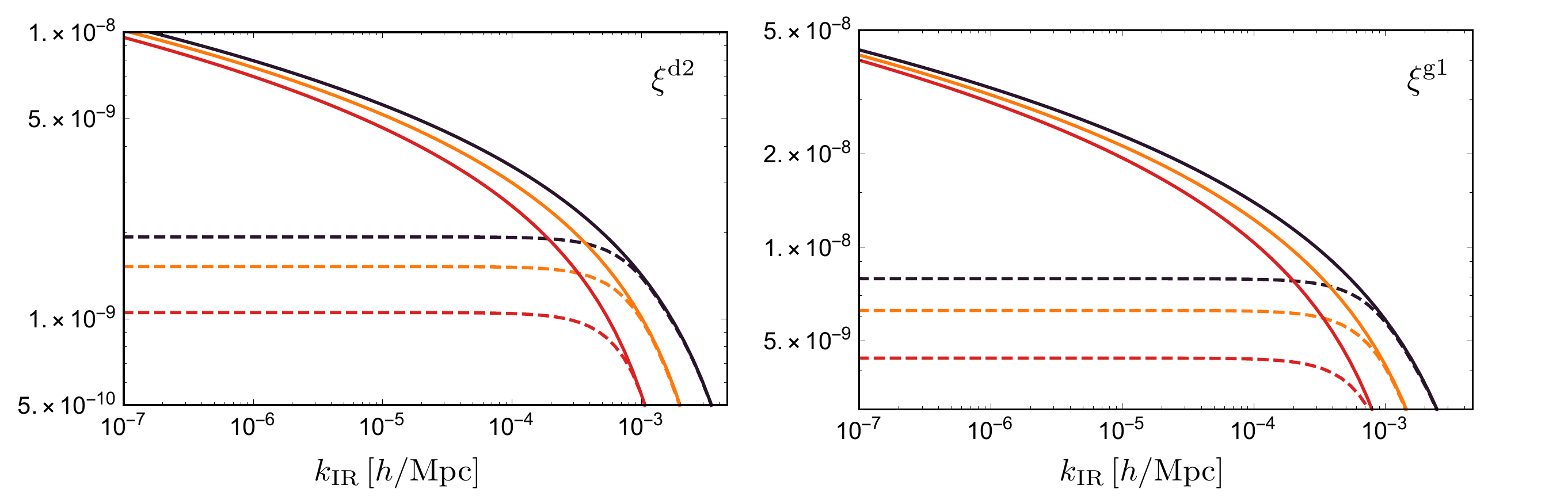}
\caption{\label{f:div1} The divergent (\emph{solid}) and convergent (\emph{dashed}) correlation function for two potential terms as a function of the IR cut-off $k_\text{IR}$. It is shown how eq.~(\ref{xireg}) regularise the Infra-Red behaviour of the correlation function. Different colours are different choices of $(r,\mu)$.}
\end{figure*}

We now turn to a problem which afflicts the 2pF contributions coming from the auto-correlation and cross-correlation of potentials terms, namely $\xi^{AB}$ with $\{A,B\} = \{\text{d2},..,\text{g5}\}$. The issue is essentially that the integral $I^4_0$ has an Infra-Red divergence. It is in fact known that the variance of the curvature  has an IR divergence \cite{Barausse:2005nf,Gerstenlauer:2011ti,Bertacca:2012tp,Biern:2016kys}:
\be
\langle \zeta^2(x) \rangle = \int_{k_\text{IR}}^\infty \frac{dk}{k} \mathcal{P}_\zeta(k) \sim \left( \frac{k_\text{IR}}{k_*}\right)^{n_s-1} \xrightarrow{k_\text{IR}\rightarrow 0} \infty  ~~  \mbox{ for } n_s\leq 1\,. \label{eq:div}
\ee
This is a problem for the integral $I^4_0$. Recall that $\Phi_k \sim \zeta_k S_\Phi (k) \sim k^{-2} \zeta_k S_D(k)$ which, with the large scale behavior $S_\Phi (k) \rightarrow 1$, gives  $S_D(k) \sim k^2$ for $k \rightarrow 0$. This implies
\be
I^4_0 (r) =  \int \frac{dk\, k^2}{2 \pi^2} \,P(k) \,\frac{ j_0( k r )}{(k r)^4} \sim \int \frac{dk}{k^5} S_D(k)^2 \mathcal{P}_\zeta(k) \sim \int \frac{dk}{k} \mathcal{P}_\zeta(k) \rightarrow \infty\,.
\ee
This means that the auto- and cross-correlations of potential terms (d2,..,g5) grow indefinitely as $k_\text{IR} \rightarrow 0$ (as they all depend on $I^4_0$, see appendix~\ref{appcoeff}). It is clear that this is an unphysical divergence as $\Delta(\bn,z)$ and its 2pF, $\xi(\theta,z_1,z_2)$, are observables and they can therefore not diverge. 

\begin{figure*}[h]
\centering
\includegraphics[scale=0.56]{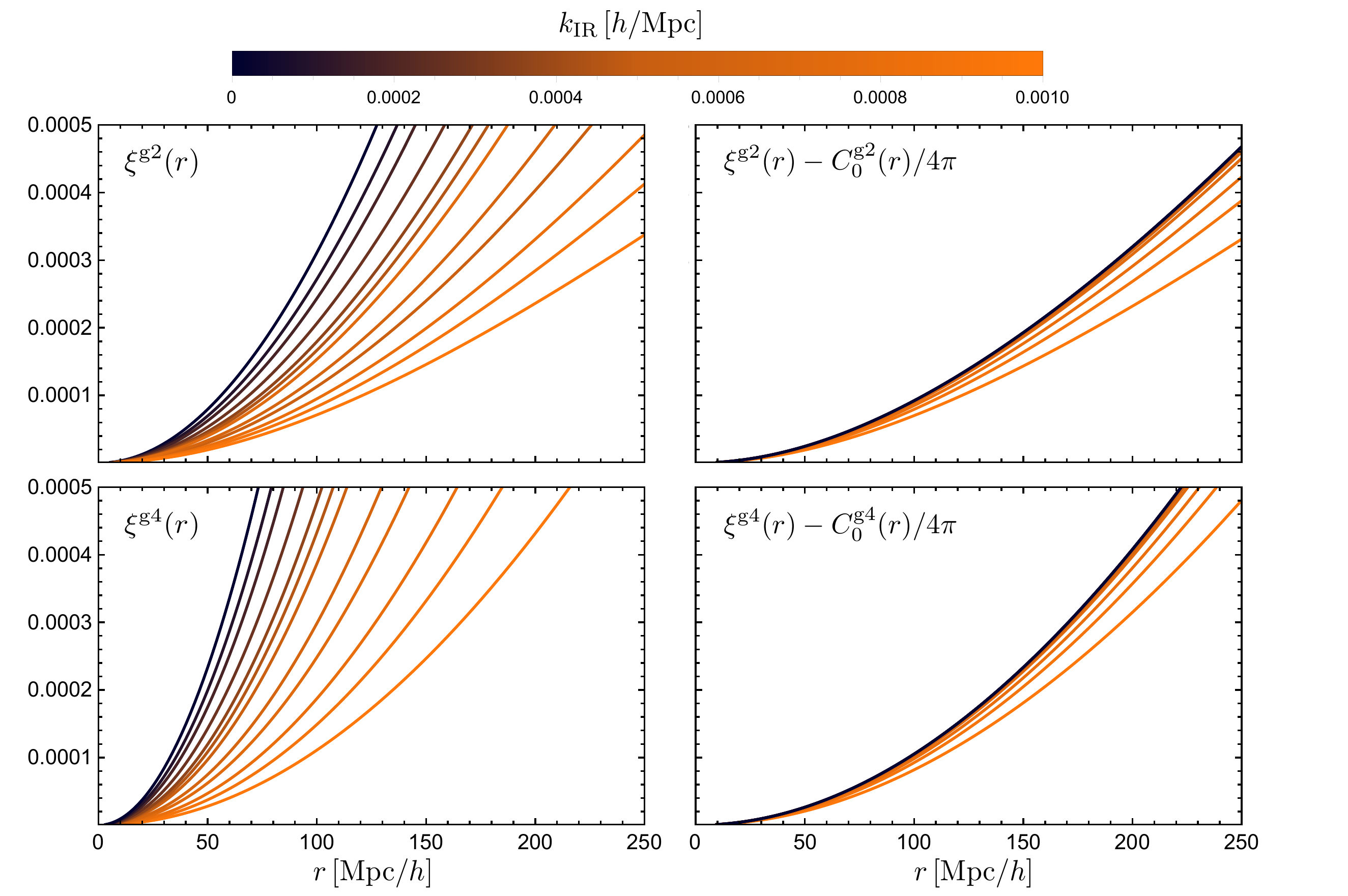}
\caption{\label{f:div2} \emph{Left}: the divergent correlation function for two potential terms as a function of separation $r$ for different values of the IR cut-off. \emph{Right}: subtracting the unobservable monopole the correlation function converges for "reasonable" values of $k_\text{IR} \lesssim 10^{-4}$. }
\end{figure*}

To understand why the divergence in~\eqref{eq:div} does not contribute to the observable, let us go back to the definition of $\De(\bn,z)$:
\be
\Delta(\bn,z) = \frac{N(\bn,z)-\langle N \rangle_\Omega(z)}{\langle N \rangle_\Omega(z)} \,,
\label{deom}
\ee
where $N(\bn,z)$ is the number of galaxies in direction $\bn$ at redshift $z$ and $\langle N \rangle_\Omega(z)$ is the directional average of $N(\bn,z)$:
\be
\langle N \rangle_\Omega(z) = \frac{1}{4\pi} \int d\Omega_\bn \, N(\bn,z)\,.
\ee
It is clear from eq.~(\ref{deom}) that $\langle \Delta \rangle_\Omega=0$. This simply reflects the fact that $\Delta$ is the departure from the average number of galaxies. Since in linear perturbation theory directional average and ensemble average commute~\cite{Bonvin:2015kea}, we also have $\langle \xi(\theta,z_1,z_2) \rangle_\Omega=0$, meaning that the correlation function does not have a monopole contribution. Physically this comes from the fact that an observer will include into $\langle N \rangle_\Omega$ not only the background but also all the IR modes which he {\it cannot} distinguish from the background. This includes super-horizon modes as well as terms at the observer, which we have neglected in this work specifically for this reason. All these modes contribute only to the monopole\footnote{The peculiar velocity at the observer also induces a dipole contribution~\cite{Yoo:2009au}, which is irrelevant for this regularisation discussion.} and as they are included in the directional average they are subtracted in eq.~\eqref{deom}, leading to $C_0=0$. If we apply these considerations to eq.~(\ref{e:xiCl}) we see that we are able to cure the divergence by explicitly removing the monopole
\be
\xi_g \longrightarrow \xi_g - C_0/4 \pi \,.
\label{xireg}
\ee
In principle this line of reasoning could be applied to all the contributions to the 2pF, however we only regularise in this way the contributions for which the monopole $C_0$ is divergent (auto- and cross-correlations of potential terms (d2,..,g5)). For the other terms this correction is negligible. Eq.~(\ref{xireg}) can be easily implemented in the code as it amounts to a redefinition of $I^4_0(r)$:
\be
I^4_0(r) \longrightarrow \frac{1}{r^4}\int \frac{dk}{2\pi^2} k^{-2} P(k)\bigg(j_0(k r)- j_0(k \chi_1) j_0(k \chi_2) \bigg) \,. 
\label{i04sub}
\ee
The result of this procedure is shown in figures~\ref{f:div1} and~\ref{f:div2}.\\

We point out that from a theoretical point of view the regularisation of the divergence can be achieved by consistently keeping track of the terms at the observer~\cite{Biern:2016kys,Biern:2017bzo}. The resulting 2pF will be gauge invariant, consistent with the equivalence principle\footnote{Inconsistency with the equivalence principle can be regarded as the reason for which the divergence arises: a term like $\De^\text{g2}$, for example, is given by the value of the gravitational potential at the source $\Phi_s$. This is not observable, while considering a counter term at the observer $\Phi_s-\Phi_o$ does not only agree with the equivalence principle but it also regularises the divergence.} and free of divergences. However to achieve this result one has to ensemble average over different realisations of the perturbation fields at the observer. This procedure leads to a result which is not linked with the observable correlation function since we can only observe from our position and therefore the ergodic theorem cannot be applied on the observer position. 

\subsection{Estimators and the covariance matrix}\label{eec}
The correlation function can be estimated in several ways from a given galaxy catalog. In this section we present the two estimators for the multipoles of the 2pF that we consider in {\sc coffe} and we compute their covariance matrix. We start by splitting a catalog covering a fraction of the sky $f_\text{sky}$ and a redshift interval  $(\bar z - \de z, \bar z +\de z)$, amounting to a total volume $V$, into pixels of comoving size $L_p$. We then count the number of galaxies $N_i$ in each pixel $i$ at redshift $z_i$ in the sky and we define $\De_i \equiv \De (\bx_i)$ as in eq.~(\ref{deom}):
\be
\De_i = \frac{N_i-\langle N \rangle_\Omega(z_i)}{\langle N \rangle_\Omega(z_i)} \,,
\ee 
where the directional average $\langle N \rangle_\Omega$ is performed over all the pixels at redshift $z_i$. The simplest estimator we can construct for the multipoles is then
\be\label{e:xihat}
\hat \xi_\ell (r,\bar z) = \beta_\ell \sum_{ij} \De_i \De_j \PP_\ell(\mu_{ij}) \delta_K(r_{ij}-r)\,,
\ee
where $r_{ij}=|\bx_i-\bx_j|$ is the distance between the two pixels and $\mu_{ij}=r_\pa/r = (\chi(z_1)-\chi(z_2))/r$ represents the orientation of the pixels. The function $\de_K$ denotes the (dimensionless) Kronecker delta 
%using the fact that in a finite catalog there appear a finite number of values $x_{ij}$. 
representing the fact that in a binned catalog the values $r_{ij}$ are discrete. The normalisation factor $\beta_\ell$ is obtained by imposing that in the continuum limit the ensemble average of the estimator satisfies $\langle \hat \xi_\ell \rangle = \xi_\ell$. One finds (more details are given in appendix~\ref{a:est})
\be
\beta_\ell = \frac{2\ell+1}{4\pi} \frac{L_p^5}{r^2 V} \,.
\ee

Equation~\eqref{e:xihat} is the estimator which is usually used in redshift surveys~\footnote{Note that in practice the Landy-Szalay estimator~\cite{1993ApJ...412...64L} is used, in order to account for the geometry of the surveys and for irregularities in the galaxy distribution.}. However to obtain this result we have made one important approximation, which is to neglect time-evolution in our redshift shell. We have indeed assumed that all pairs of pixels $(i,j)$ in eq.~\eqref{e:xihat} have the same mean redshift $\bar z$. It is only under this assumptions that the estimator~\eqref{e:xihat} is unbiased, i.e. that $\langle \hat \xi_\ell \rangle = \xi_\ell$. In practice however, we know that the galaxy distribution evolves with redshift, so that each pair of pixels $(i,j)$ contributes in a slightly different way to the sum. This is especially relevant when computing the multipoles of the correlation function at large separation $r$, for which thick redshift bins must be used~\footnote{In order to measure the multipoles at large separation $r$, we need indeed a redshift bin thicker than $r$ in order to include pairs with all orientations in the average over $\mu$.}. In this case, the mean of~\eqref{e:xihat} can be different from the theoretical predictions $\xi_\ell$ and the estimator is therefore biased. For this reason we propose a second estimator, which distinguishes between different mean redshifts $z_{ij}=(z_i+z_j)/2$ inside the redshift bin, and which is therefore unbiased also in the full sky regime
\be\label{e:Xi}
\hat \Xi_\ell (r,\bar z ,\de z) = \gamma_\ell \sum\limits_{\{z_k\}} W(z_k) \sum\limits_{i,j} \frac{1+\cos\theta_{ij}}{2r_j^2} \De_i \De_j\, \mathcal{P}_\ell(\mu_{ij})\, \de_K(r_{ij}-r)\de_K(z_{ij}-z_k) \,,
\ee
which sums all the pairs at fixed separation $r$ and at fixed mean redshift $z_k$ and then sums over all the different redshifts in the bin $\{ z_k\}$ so that no pair in the catalog is lost.  Here $\theta_{ij}$ is the angle between $\bx_i$ and $\bx_j$ (note that this is not $\mu_{ij} = 2(|\bx_i|-|\bx_j|)/ (|\bx_i|+|\bx_j|)$. The expectation value of this new estimator is the quantity $\Xi_\ell$ (which we can compute with {\sc coffe}) defined as
\be\label{e:Xicont}
\Xi_\ell (r,\bar z ,\de z) = \HH_0\int\limits_{\bar z-\de z}^{\bar z +\de z} dz \, \frac{W(z)}{\HH(z)(1+z)} \xi_\ell(r,z) \,.
\ee
Here $W(z)$ denotes the redshift distribution (normalised to unity) and the normalisation factor $\gamma_\ell$ has to be chosen as
\be
\gamma_\ell = \frac{2\ell+1}{(4\pi)^2} \frac{L_p^5\HH_0}{r^2 f_\text{sky}} \,.
\ee
Details of the derivation of this result are given in appendix~\ref{a:est}, where we show that $\hat \Xi_\ell$ is an unbiased estimator of $\Xi_\ell$ at all separations. Note that the non-trivial factor 
\be
\frac{1+\cos\theta_{ij}}{2r_j^2}\, ,
\ee 
that we have introduced in eq.~\eqref{e:Xi} is necessary to find eq.~\eqref{e:Xicont}. It accounts for the geometry of the average over pairs in the full-sky regime. It can be expanded as 
\be
\frac{1+\cos\theta_{ij}}{2r_j^2} \simeq \frac{1}{\bar \chi^2} \left(1 \pm \mu \frac{r}{\bar \chi}  +O\left(\frac{r}{\bar\chi} \right)^2\right)\,,
\label{expcos}
\ee
and it reduces to $1/\bar \chi^2$ in the flat-sky approximation.

Let us finally address one important subtlety in the definition of eq.~(\ref{e:Xicont}). Equation~\eqref{e:Xicont} is a weighted average of $\xi_\ell(r, z)$, with $z$ running over the size of the redshift bin, i.e. from $z(\chi_1)=\bar z-\delta z$ to $z(\chi_2)=\bar z+\delta z$. In practice however, the mean redshift of the pair of galaxies $z$ cannot take all the values between $z(\chi_1)$ and $z(\chi_2)$. The calculation of the multipoles defined in eq.~\eqref{multdef} contains indeed a sum over all orientations $\mu$. However, for a given separation $r$ and orientation $\mu$ not all values of $z$ are permitted. More precisely, the allowed values are $z \in [z(\chi_1 +r/2), z(\chi_2-r/2)]$.
If we want to take care of this subtlety theoretically, we have to make the limits of integration and the redshift distribution $W(z)$ in eq.~(\ref{e:Xicont}) $r$-dependent. For a simple top-hat distribution, {\sc coffe} computes
\be\label{e:Xicont3}
\Xi_\ell (r,z_1 ,z_2) = \frac{\HH_0}{z_2(r)-z_1(r)}\int\limits_{z_1(r)}^{z_2(r)} dz \, \frac{\xi_\ell(r,z) }{\HH(z)(1+z)} \,,
\ee
where $z_1(r)=z(\chi_1+ r/2)$ and $z_2(r)=z(\chi_2- r/2)$.

\begin{figure*}[t]
\centering
\includegraphics[scale=0.76]{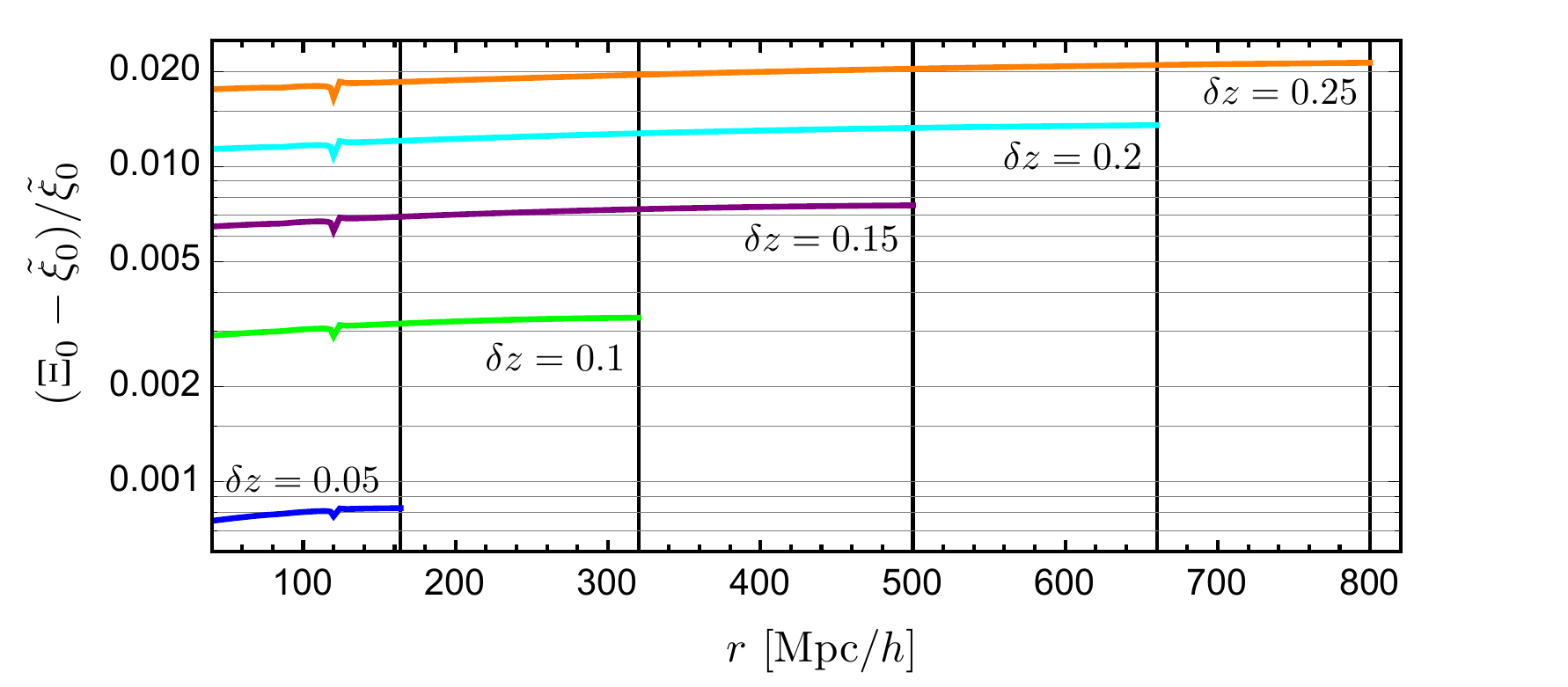}
\caption{\label{f:deltax} The fractional difference $(\Xi_\ell (r) -\tilde{\xi}_\ell(r))/\tilde{\xi}_\ell(r)$ for the monopole $\ell =0$ at redshift $\bar z =1$. The $\Xi_\ell$ are computed in redshift bins with different half-widths $\delta z$. The monopole at $\bar z$ can only be calculated out to $r\sim 2\de z/H(z)$. The 'glitch' at $r\simeq 120 \, \text{Mpc}/h$ comes from the monopole going through zero. The result for $\ell=2,4$ is similar.}
\end{figure*}

We can now compare our two estimators. In figure~\ref{f:deltax} we show the fractional difference, at $\bar z =1$, between the mean of the two estimators: $\xi_\ell(r)$ and $\Xi_\ell(r,\delta z)$  for different values of the half-width of the bin $\delta z$. The main difference between the two estimators is a different normalisation: this is because the $\xi_\ell$ are computed exactly at $\bar z =1$, while the $\Xi_\ell$ are averaged over the redshift bin. The wider the bin, the larger is the deviation from the multipole at the mean redshift. The second difference is more fundamental since it is directly due to the evolution of the galaxy number counts with redshift. This effect is slightly scale-dependent and it can be isolated in the following way: let us define a flat-sky $\Xi_\ell$ starting from the flat-sky $\xi_\ell$. In this case, as evolution is neglected in the flat-sky limit, the only difference between the two estimators would be due to their different normalisation. In particular, in the flat-sky limit we can separate the $z$- and $r$-dependence of the multipoles, as in eq.~(\ref{fsxi}), to obtain
\be
\Xi_{\ell}(\bar z)_\text{flat-sky} = \left(\frac{\HH_0}{2 \delta z}\frac{1}{c_\ell(\bar z) D_1^2(\bar z)}\int\limits_{\bar z - \delta z}^{\bar z +\delta z} dz \, \frac{c_\ell(z) D_1^2(z)}{\HH(z)(1+z)} \right)\xi_{\ell}(\bar z)_\text{flat-sky} \,,
\ee
where the $c_\ell$'s are defined in eqs.~\eqref{e:c0}-\eqref{e:c4}. In the full-sky regime, at large separations, we expect a deviation from this simple behaviour. In figure~\ref{f:deltax} we therefore normalise the multipoles $\xi_\ell$ with
\be
\tilde{\xi}_\ell \equiv \left( \frac{\HH_0}{2 \delta z}\int\limits_{\bar z - \delta z}^{\bar z +\delta z} dz \, \frac{1}{\HH(z)(1+z)} \right) \xi_\ell\,.
\ee
In this way, we get rid of the difference due to the normalisation and we show only the intrinsic difference due to evolution. Overall the difference between the estimators is small, but it can be substantial (of order $1\%$) if a thick redshift bin is considered. Finally, let us emphasise again that whereas $\Xi_\ell$ is an unbiased estimator of $\hat \Xi_\ell$ at all separations, $\xi_\ell$ is biased at large separations due to evolution. The order of magnitude of this bias is related to the difference plotted in figure~\ref{f:deltax}. For very thick redshift bins, $\xi_\ell$ is therefore not a reliable estimator of the multipoles and $\Xi_\ell$ should be used instead.

We can now compute the covariance matrix for the two estimators:
\bea
&\text{cov}_{\ell\ell'}^{(\xi)}(r,r') \equiv  \Big\langle \hat \xi_\ell (r) \hat \xi_{\ell'}(r') \Big\rangle- \Big\langle  \hat \xi_\ell (r)\Big\rangle \Big\langle\hat \xi_{\ell'}(r') \Big\rangle \,,\\
&\text{cov}_{\ell\ell'}^{(\Xi)}(r,r') \equiv  \Big\langle \hat \Xi_\ell (r) \hat \Xi_{\ell'}(r') \Big\rangle- \Big\langle  \hat \Xi_\ell (r)\Big\rangle \Big\langle\hat \Xi_{\ell'}(r') \Big\rangle \,.
\eea
The variance of the number counts has two contributions
\be
\langle \De_i\De_j\rangle = \frac{1}{d\bar N} \de_{ij} + C_{ij} \,.
\ee
The first term accounts for shot noise, where $d\bar N$ is the average number of tracers per pixel. It comes from the fact that we Poisson sample from the underlying smooth density distribution. Shot noise contributes only to the correlation function at zero separation, i.e. when $i=j$. The second term is the cosmic variance contribution. For simplicity we perform the covariance calculation in the flat-sky approximation (which means we stop at the $0^\text{th}$-order term in eq.~(\ref{expcos})) and we consider only the density and redshift-space distortion contributions. Since this is by far the dominant term, it is a good approximation to the full result. Assuming Gaussianity (i.e. we write 4-point functions as products of 2-point functions) and following the procedure outlined in~\cite{Bonvin:2015kuc,Hall:2016bmm}, we can express the covariance matrix in terms of Wigner's  3j-symbols as
\be
\begin{split}
\text{cov}_{\ell\ell'}^{(\xi)}(r_i,r_j) = \frac{ i^{\ell-\ell'} }{V} &\Bigg[  \frac{2\ell+1}{2\pi \bar n^2 L_p r^2} \de_{ij} \de_{\ell\ell'}+ \frac{1}{\bar n} \mathcal{G}_{\ell\ell'} (r_i,r_j,\bar z) \sum\limits_\sigma c_\sigma\tj{\ell}{\ell'}{\sigma}{0}{0}{0}^2\\
&+ \mathcal{D}_{\ell\ell'} (r_i,r_j,\bar z) \sum\limits_\sigma \tilde c_\sigma\tj{\ell}{\ell'}{\sigma}{0}{0}{0}^2 \Bigg] \,,
\end{split}
\label{covxi}
\ee
\be
\begin{split}
\text{cov}_{\ell\ell'}^{(\Xi)}(r_i,r_j) &= \frac{ i^{\ell-\ell'} }{4\pi f_\text{sky}}  \int\limits_{\bar z-\de z}^{\bar z +\de z} dz \, \frac{W^2(z)}{\HH(z)\chi^2(z)(1+z)} \Bigg[  \frac{2\ell+1}{2\pi \bar n^2 L_p r^2} \de_{ij} \de_{\ell\ell'} \\& +\frac{1}{\bar n} \mathcal{G}_{\ell\ell'} (r_i,r_j,z) \sum\limits_\sigma c_\sigma\tj{\ell}{\ell'}{\sigma}{0}{0}{0}^2 +\mathcal{D}_{\ell\ell',z} (r_i,r_j,z) \sum\limits_\sigma \tilde c_\sigma\tj{\ell}{\ell'}{\sigma}{0}{0}{0}^2 \Bigg] \,,
\end{split}
\label{covXi}
\ee
where $\bar n$ is the mean number density\footnote{In the covariance we ignore the redshift dependence of $\bar n$ and set $\bar n \equiv n(\bar z)$ inside a given redshift bin. A method to include $\bar n (z)$ consistently can be found in~\cite{2013MNRAS.431.2834X}.} in the redshift bin and we have defined
\begin{align}
&\mathcal{G}_{\ell\ell'} (r,r',z) = \frac{2 (2\ell+1)(2\ell'+1)}{\pi^2} \int dk \, k^2  P(k,z) j_\ell( k r) j_{\ell'}(k r') \,, \label{e:Gll}\\
& \mathcal{D}_{\ell\ell'} (r,r',z) = \frac{(2\ell+1)(2\ell'+1)}{\pi^2} \int dk \, k^2  P^2(k, z) j_\ell( k r) j_{\ell'}(k r') \,, \label{e:Dll}
\end{align}
together with the modified coefficients
\begin{align}
&\tilde  c_0 = c_0^2 +\frac{c_2^2}{5}+\frac{c_4^2}{9} \,, \label{e:c0tilde}\\ \displaybreak[0]
&\tilde  c_2 = \frac{2}{7}c_2 (7c_0+c_2) +\frac{4}{7} c_2 c_4+\frac{100}{693}c_4^2 \,, \\ 
&\tilde  c_4 = \frac{18}{35} c_2^2 +2 c_0 c_4 +\frac{40}{77} c_2 c_4 +\frac{162}{1001} c_4^2 \,,\\
&\tilde  c_6 = \frac{10}{99} c_4 (9 c_2 +2 c_4) \,,\\
&\tilde  c_8 = \frac{490}{1287} c_4^2 \,. \label{e:c8}
\end{align}
These results are also derived in appendix~\ref{a:est}, while in fig.~\ref{f:covimg} we show the covariance matrix for the monopole, the quadrupole and their cross-correlation.

\begin{figure*}[ht]
\centering
\includegraphics[scale=0.51]{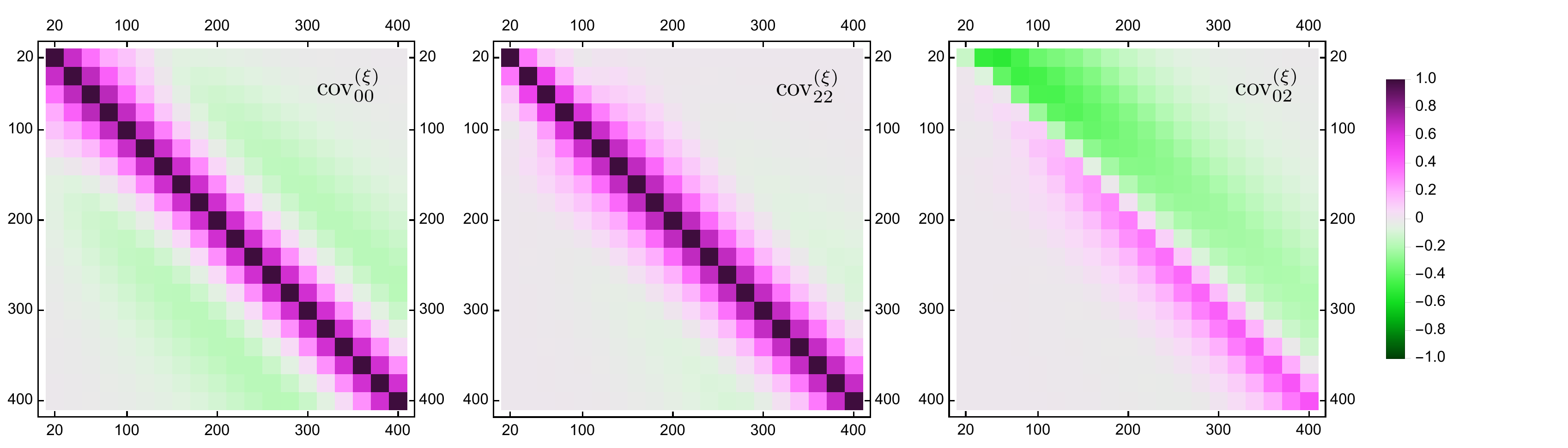}
\caption{\label{f:covimg} The covariance matrix for the monopole, the quadrupole and their cross-correlation, normalised as $\text{cov}_{\ell\ell',ij}/(\text{cov}_{\ell\ell,ii}\text{cov}_{\ell'\ell',ii}\text{cov}_{\ell\ell,jj}\text{cov}_{\ell'\ell',jj})^{1/4}$. SKA2 specifications are used here and we plot the covariance for the middle bin of the 5-bin configuration, i.e. $L_p= 20 \, \text{Mpc}/h$.}
\end{figure*}

%%%%%%%%%%%%%%%%%%%%%%%%%%%%%%%%%%Simple application%%%%%%%%%%%%%%%%%%%%%%%%%%%%%%%%%%%%%%%%%%%%

\section{A simple application: is lensing detectable?}\label{application}
\begin{figure*}[ht]
\centering
\includegraphics[scale=0.55]{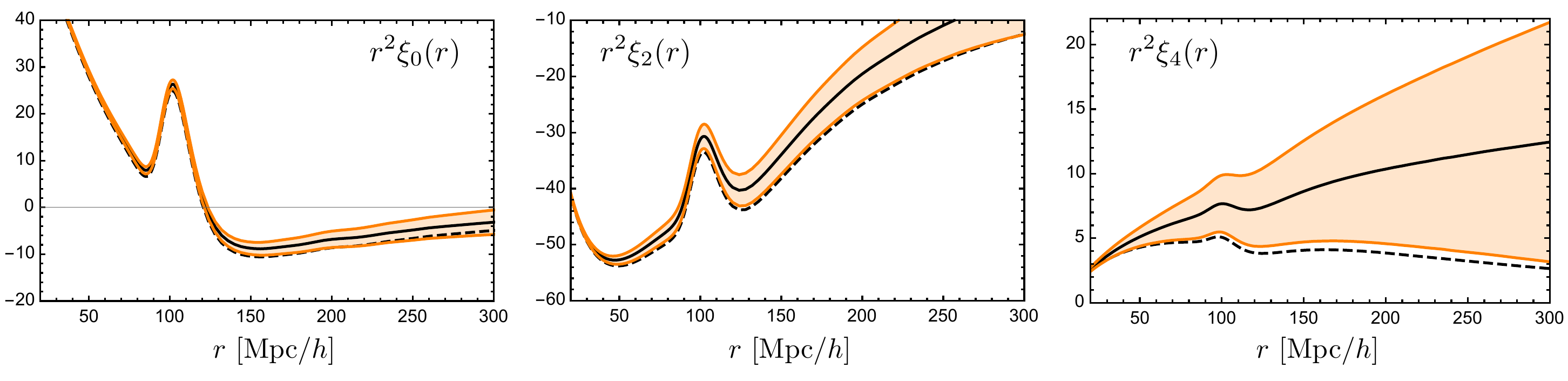}
\caption{\label{f:multlen} The monopole ({\it left}), the quadrupole ({\it middle}) and the hexadecapole ({\it right}) at $\bar z =1.5$. Solid lines have $A_L=1$ while dashed lines have no lensing contribution $A_L=0$. Specifications, in particular biases, are for an SKA2-like survey.}
\end{figure*}

As a first, simple application of {\sc coffe} we want to discuss the feasibility of measuring the lensing contribution in the correlation function with future galaxy surveys. In order to do so, we introduce an artificial parameter $A_L$, encoding the amplitude of the lensing signal, in the multipoles of the two-point function. Schematically, with the notation of eq.~(\ref{notxi}) we write (neglecting the Doppler and potential terms)
\be
\xi_\ell = \xi_\ell^{\text{st}} + A_L\,  \xi_\ell^L \,,
\ee
where
\be
\xi_\ell^{\text{st}} =  \langle \text{den}+\text{den} \rangle_\ell + \langle \text{den}+\text{rsd} \rangle_\ell + \langle \text{rsd}+\text{den} \rangle_\ell + \langle \text{rsd}+\text{rsd} \rangle_\ell \,,
\ee
represent the standard density and redshift-space distortion term, and
\be
 \xi_\ell^L = \langle \text{den}+\text{len} \rangle_\ell + \langle \text{len}+\text{den} \rangle_\ell+ \langle \text{rsd}+\text{len} \rangle_\ell+ \langle \text{len}+\text{rsd} \rangle_\ell + \langle \text{len}+\text{len} \rangle_\ell \,,
\ee
is the lensing contribution. Clearly the physical value of the lensing amplitude is $A_L=1$ and we want to forecast the precision with which we can measure it. In figure~\ref{f:multlen}, we show the monopole, quadrupole and hexadecapole with ($A_L=1$) and without ($A_L=0$) the lensing contribution. The shadowed regions show the size of the error-bars for an SKA2-like survey (specifications given below).

The Fisher matrix is defined as 
\be\label{Fisher}
F_{\alpha\alpha'} \equiv \frac{\pd^2 \chi^2}{\pd \alpha \pd \alpha'} = \sum\limits_{\ell,\ell',i,j} \frac{\pd \langle \hat \xi_{\ell} \rangle(r_i)}{\pd \alpha} \bigg|_\text{f} \text{cov}^{-1}_{\ell\ell'}(r_i,r_j)  \frac{\pd \langle \hat \xi_{\ell'}\rangle(r_j)}{\pd \alpha'} \bigg|_\text{f}\, ,
\ee
where $\alpha$ and $\alpha'$ are the parameters we want to constrain and $ |_\text{f}$ means evaluation at some fiducial values of the parameters. The sum runs over all pixels' separations $r_i, r_j$ in the survey as well as over the even multipoles $\ell, \ell'=0, 2, 4, 6$. Note that the covariance matrices account for both correlations between different pixels' separations, $r_i\neq r_j$, and correlations between different multipoles, $\ell\neq \ell'$. The Cram\'er-Rao bound states that we can assign the $1\sigma$ uncertainty as
\be
\sigma_\alpha = \sqrt{(F^{-1})_{\alpha\alpha}}\,,
\ee
and this gives the smallest possible achievable error on $\alpha$. We assume that the thickness of the redshift bins in which we split our catalog is big enough so that we can treat them as uncorrelated, implying
\be
F_{\alpha\alpha'}^{\text{tot}} = \sum\limits_{\{\bar z_i \}} F_{\alpha\alpha'}(\bar z_i) \,.
\ee 
Furthermore, for simplicity, we consider only the parameter $A_L$ and, instead of marginalizing over the remaining cosmological parameters, we fix them to: $\Omega_\text{cdm} = 0.26$, $\Omega_b = 0.048$, $h=0.676$, $A_s = 2.22 \times 10^{-9}$ and $n_s = 0.96$. In this case we only have
\be\label{FisherAA}
\begin{split}
F_{A_L A_L} \equiv F &= \sum\limits_{\ell,\ell',i,j} \frac{\pd \langle \hat \xi_{\ell} \rangle(r_i)}{\pd A_L} \bigg|_\text{f} \text{cov}^{-1}_{\ell\ell'}(r_i,r_j)  \frac{\pd \langle \hat \xi_{\ell'}\rangle(r_j)}{\pd A_L} \bigg|_\text{f}  \\
& = \sum\limits_{\ell,\ell',i,j}  \xi_\ell^L(r_i) \bigg|_\text{f} \text{cov}^{-1}_{\ell\ell'}(r_i,r_j)  \xi_\ell^L(r_j) \bigg|_\text{f} \,.
\end{split}
\ee
Note that the parameter $A_L$ does not have a direct physical interpretation; however, it allows us to estimate the signal-to-noise (S/N), a measure of the sensitivity to the lensing signal in galaxy clustering. This is an important information, especially as a high S/N is needed to test deviations from general relativity, for example with the widely used $(\Sigma,\mu)$ parametrization (see e.g. \cite{Gubitosi:2012hu,Simpson:2012ra,Ferte:2017bpf} and references therein): $\mu \ne 1$ represents a modification to Poisson equation while $\Sigma \ne 1$ represents a modification to the gravitational slip relation (see eqs.~(\ref{modpoiss}),(\ref{modspli})). The standard terms constrain $\mu$, while one needs to be sensitive to the lensing potential to constrain also $\Sigma$. \\

In this analysis we make one optimistic assumption and one conservative assumption. The optimistic one, as we mentioned, is to neglect the parameter degeneracies that will increase the actually achievable error bar on $A_L$. A forecast study on all the cosmological parameters is left as future work~\cite{prep}. The conservative one is to treat the lensing term within linear perturbation theory, while non-linearities increase the lensing signal \cite{Tansella:2017rpi}: {\sc coffe} is for the moment a fully linear code and, as we discuss it in section~\ref{next}, pushing its capabilities beyond the linear treatment is amongst our priorities.  \\

The forecast~(\ref{FisherAA}) is easily done with {\sc coffe} and we compute the result for the signal-to-noise, which is simply given by
\be
\rm{S/N} =\sqrt{F} = 1/\sigma_{A_L} \,,
\label{sndef}
\ee
with $A_L|_\text{f} =1$.\\

\begin{figure*}[ht]
\centering
\includegraphics[scale=0.51]{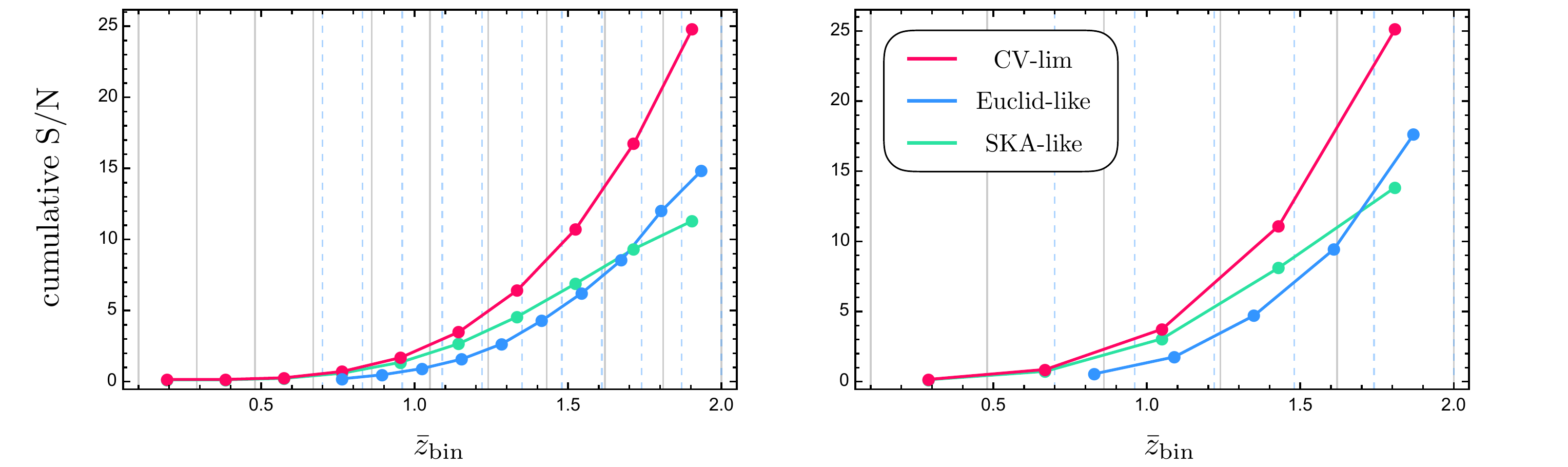}
\caption{\label{f:fora} The cumulative signal-to-noise on the parameter $A_L$ for three different survey specifications and two different choice of binning, as explained in the text.}
\end{figure*}

In Figure~\ref{f:fora} we show the results for the signal-to-noise for three different spectroscopic survey specifications: an Euclid-like survey (specifications given in \cite{2011arXiv1110.3193L}), an SKA2-like survey (specifications given in \cite{Villa:2017yfg}) and a survey limited only by cosmic variance, in which shot noise is neglected (essentially performing the limit $\bar n \rightarrow \infty$ in eq.~(\ref{covxi})). We split the surveys into 5 bins (right panel of figure~\ref{f:fora}) or 10 bins (left panel of figure~\ref{f:fora}) to accommodate the full redshift range: $z \in [0.1,2.0]$ for the SKA2 and the CV-limited survey and $z \in [0.7,2.0]$ for Euclid (respectively solid and dashed vertical lines in fig.~\ref{f:fora}). For the 5-bins configuration we chose $L_p = 20 \, \text{Mpc}/h$ while for the 10-bins configuration we set $L_p = 10 \, \text{Mpc}/h$. We include separations from $r_{\rm min}=L_p$ to $r_{\rm max}=\chi(\bar z_{\rm bin}+\delta z)-\chi(\bar z_{\rm bin}-\delta z)$.
The redshifts $\bar z<1$ contribute very little to the signal. Only for $\bar z>1$ sufficient lensing has accumulated to be truly visible in the correlation function. The S/N for the 5-bin configuration is somewhat larger than the one for the 10 bin. This is due to the fact that lensing dominates for large radial separation. Since we neglect correlations between different bins in the calculation of the Fisher matrix, we include more correlations at large separations when we have 5 bins than when we have 10 bins. The results  for the cumulative S/N of eq.~(\ref{sndef}) on the lensing amplitude $A_L$ are summarised as follows: for the 5-bins splitting we obtain
\be
\begin{split}
&\rm{S/N} \big|_\text{CV-lim} \simeq 25.2 \,, \\
&\rm{S/N} \big|_\text{Euclid-} \simeq 17.6 \,, \\
&\rm{S/N} \big|_\text{SKA} \simeq 13.8 \,,
\end{split}
\ee
while for the 10-bins splitting they are slightly lower. Note that the monopole alone contributes to $\sim 45 \%$ of the total $S/N$, the quadrupole to $\sim 30\%$, the hexadecapole to $\sim 15\%$ while, interestingly, the $\ell=6$ multipole contributes to roughly $10\%$ of the total signal-to-noise. This can be explained as the balance between two different effects: on the one hand the lensing contribution is more relevant for higher multipoles, but on the other hand, as seen in fig.~\ref{f:multlen}, the covariance also gets bigger for higher $\ell$. We therefore conclude that upcoming galaxy surveys will be able to detect the lensing signal in the 2pF. This will open the possibility to put constraints in the $(\Sigma,\mu)$ plane from the clustering signal alone. \\

Note that here we have used the $\xi_\ell$ for the forecasts. Since we split the survey in bins of half-width $\delta z = 0.05$ and $\delta z =0.1$, given the discussion in section \ref{eec}, we do not expect these results to change if we use the $\Xi_\ell$ instead.

%%%%%%%%%%%%%%%%%%%%%%%%%%%%%%%%%%Structure of the code%%%%%%%%%%%%%%%%%%%%%%%%%%%%%%%%%%%%%%%%%%%%

\section{Structure of the code}\label{struct}
{\sc coffe} is entirely written in \code C and the code can be divided into several key \source{struct}s and functions.
The main flow of the program can be summarized as follows:
\begin{enumerate}
\item Read the settings file containing all of the necessary parameters: cosmological parameters, the input $P(k)$ and desired output. 
\item Compute and store the background quantities.
\item Calculate all of the $I_\ell^n(r)$ using an implementation of the \code{2-FAST} algorithm for a fixed number of separations, specified by the user;  $I^4_0$ is only computed if one of the following contributions is requested: d2, g1, g2, g3, g4, g5.
\item Compute one of the following, depending on input:
\begin{itemize}
\item angular correlation function $\xi(\theta,\bar z)$
\item full sky correlation function $\xi(\mu,r,\bar z)$
\item multipoles of the correlation function $\xi_\ell(r, \bar z)$
\item redshift averaged multipoles $\Xi_\ell (r,\bar z ,\de z)$
\item covariance of multipoles $\text{cov}_{\ell\ell'}^{(\xi)}$
\item covariance of redshift averaged multipoles $\text{cov}_{\ell\ell'}^{(\Xi)}$
\end{itemize}
\item Save the necessary output.
\item Perform a memory cleanup and exit.
\end{enumerate}
In the next sections we will go over the structure in more detail. For more information, the interested reader can consult the user manual, available at \\\url{https://cosmology.unige.ch/content/coffe} or at \url{https://github.com/JCGoran/coffe}. The manual also contains detailed instructions on how to run the code.

\subsection{The parser \& background modules}
The parser module is used for the parsing of the structured settings file and making sure all of the values in the input are valid.
The library used for parsing is the \code{libconfig} library\footnote{described in \url{https://hyperrealm.github.io/libconfig/}}.\\

The background module is responsible for calculating all of the derived redshift dependent quantities in a $\Lambda$CDM or $w$CDM cosmology with zero curvature, such as the Hubble rate $H(z)$, growth rate $f(z)$, comoving distance $\chi(z)$, etc. 
All of the quantities are computed at equally spaced intervals up to redshift $z=30$, with a user defined sampling rate, and are stored in \source{interpolation} structures. We have compared our results with {\sc class} and have found an agreement of order $10^{-4}$.

\subsection{The integrals $I_\ell^n(r)$}
For the integrals $I_\ell^n(r)$ we have created a native implementation of the \code{2-FAST} algorithm introduced in~\cite{Gebhardt:2017chz}.
We have tested it against the original implementation in \textbf{julia}, and have found a negligible discrepancy of the order $10^{-5}$.
As they are computed only for a discrete number of points, we again use interpolation to find their values for arbitrary $r$.\\

Note that for the non-integrated terms we have to deal with the $r\rightarrow 0$ limit of the $I^n_\ell(r)$ only if the value of the correlation function at zero separation is required. On the other hand the expressions for the integrated terms contain integrals of the type (see appendix~\ref{appcoeff})
\be
\int\limits_0^{\chi_1} d\la  \, [...] \, I^n_\ell(r) \,, \qquad \quad \int\limits_0^{\chi_1} d\la \int\limits_0^{\chi_2} d\la' \, [...] \, I^n_\ell(r) \,,
\label{SDint}
\ee
where inside the integrals we have, respectively,
\bea
r&= \sqrt{\la^2+\chi_2^2-2 \la \chi_2 \cos\theta} \,, \label{er1} \\
r &= \sqrt{\la^2+ \la'^2 -2 \la \la' \cos\theta} \,. \label{er2}
\eea
When we compute the 2pF along the line-of-sight (i.e. $\mu=1$ or $\theta=0$) the integrand evaluates at $r=0$ and hence we have to deal with this limit. We treat the following three cases separately:
\begin{enumerate}
\item For $\boldsymbol{\ell < n}$ we have $\lim_{r\to0} I^n_\ell(r) = \infty$; however, these terms always appear in the correlation function multiplied by the appropriate power $r^{n-\ell}$, so that
\be
\lim_{r\to0} r^{n-\ell} I^n_\ell(r) \sim \int \frac{dk \, k^2}{2\pi^2} P(k) k^{\ell-n} \,.
\label{Pint}
\ee
Assuming that the linear power spectrum behaves in the IR and UV as
\be 
   P(k)\sim\left\{
                \begin{array}{ll}
                  k^{n_1} \quad \quad\text{for}\,\, k\ll1\\
                  k^{n_2-4} \quad\text{for}\,\, k\gg1
                \end{array}
              \right.
\ee
the condition for which the integral in eq.~(\ref{Pint}) converges is $-3-n_1 < \ell-n < 1-n_2$, which is always satisfied for $\Lambda$CDM cosmologies\footnote{The values obtained for the linear $P(k)$ used in the figures of this paper are $n_1\simeq n_s \simeq 0.96$ and $n_2 \simeq 1$.} for the values of $\ell,n$ needed (except for $I^4_0$ in the IR as we discussed in section~\ref{IRdiv}). We use {\sc 2-fast}  to interpolate $r^{n-\ell} I^n_\ell$ until a small separation $r_\text{min} \simeq 1 \, \text{Mpc}/h$. As the {\sc 2-fast} algorithm cannot be pushed to $r\rightarrow 0$ we use the standard {\sc GSL} integrator for  $r \lesssim 1 \, \text{Mpc}/h$, where the very oscillatory behaviour of the integrands is less pronounced, and {\sc GSL} gives a reliable result.
\item For $\boldsymbol{\ell = n}$ we have $\lim_{r\to0} I^n_\ell(r) = \text{const}$, and we can simply switch to {\sc GSL} from  $r \lesssim 1 \, \text{Mpc}/h$ to  $I^n_\ell(r=0)$.
\item For $\boldsymbol{\ell > n}$ the limit gives $\lim_{r\to0} I^n_\ell(r) = 0$ and the $I^n_\ell$ go to zero as $r^{\ell-n}$. The behaviour close to $r=0$ is again captured with the {\sc GSL} integrator. Note however that capturing the overall behaviour would be really important only if the coefficients by which the  $I^n_\ell$ are multiplied diverge at zero like $r^{n-\ell}$. As all the coefficients are well behaved at zero, our procedure causes no concern.
\end{enumerate}

The $I^4_0(r)$ integral of eq.~(\ref{i04sub}) is not in a form suitable for the \code{2-FAST} algorithm and it is therefore integrated using standard \code{GSL} integration for a predefined number of separations, and then interpolated.
In this case, we have the further complication that we also need the counter term to renormalise  $I^4_0(r)$ which is a function of both comoving distances $\chi_1$ and $\chi_2$.
This term is calculated for a fixed number of points (200$\times$200) and is then 2D interpolated for all other points.

\subsection{Outputs}
To calculate the correlation function $\xi$, its multipoles $\xi_\ell$ and the redshift averaged multipoles $\Xi_\ell$, we use the $X^n_\ell$ and $Z^n_\ell$ coefficients defined in appendix \ref{appcoeff} and build the desired quantities using Eqs.~(\ref{xiAB}),(\ref{IxiAB}),(\ref{multdef}) and (\ref{e:Xicont}) respectively. The $I^n_\ell$ integrals are computed with our implementation of the \code{2-FAST} algorithm.
Integrated terms have a structure as in eq.~(\ref{SDint}). To compute them, depending on the number of integrations required, we use either standard \code{GSL} integration ($1$ integration) or one of the following options:
\begin{itemize}
\item \code{GSL} Monte Carlo methods, using either importance sampling or stratified sampling
\item the \code{CUBA} library \cite{Hahn:2004fe}, using a deterministic integrator employing cubature rules
\end{itemize}
The user can select which one to use at compile time, as well as the number of iterations at run time.\\

The covariance is built from eqs.~(\ref{covxi}),(\ref{covXi}). The challenging part of the computation are clearly the integrals $\mathcal{D}_{\ell\ell'}$ and $\mathcal{G}_{\ell\ell'}$. As the \code{2-FAST} algorithm is not optimised to compute covariances\footnote{To be precise, \code{2-FAST} allows for the computation of integrals with two Bessel functions such as $\mathcal{D}_{\ell\ell'}$ and $\mathcal{G}_{\ell\ell'}$. However the algorithm is structured to output them for a list of $r_i$ but fixed $R= r_j/r_i$. In the covariance we however need $N_p^2$ pairs of $(r_i,r_j)$, where $N_p = r_\text{max}/L_p$ is the number of pixels in the covariance. To get them, with no modification of the algorithm, we need to run \code{2-FAST} $N_p^2$ times, with a runtime not suitable for a public code. The covariance in section \ref{application} is nevertheless computed in this way: with our implementation of the  \code{2-FAST} double-Bessel algorithm (we use the main principle of the algorithm, i.e. an Hankel transform of the integrand, but not their specific hyper geometric function ${}_2F_1$ implementation).} it is (at the moment) too slow to be implemented in the public version of the code. We therefore choose to release {\sc coffe} \source{v.1.0} with the covariance implemented in GSL, which is much faster but less precise. Note that this trade of precision for speed has sometimes important drawbacks: for thick redshift bins the GSL covariance might not be positive definite because of numerical fluctuations. For this reason the results reported in section \ref{application} have been obtained with the \code{2-FAST} algorithm: in future versions of  {\sc coffe} we will optimize this for covariance calculation and release it to the public.\\

For reference, in table~\ref{Ttime} we list the run time of {\sc coffe} for the different possible outputs.

\renewcommand{\arraystretch}{1.25}
\begin{table}
\centering
\begin{tabular}{| l l l | c | }
\hline
\textbf{output} & & &\textbf{time} \\
  \hline			
  $\xi(r,\mu)$ &(den+rsd) &up to $\sim 1000 \, \text{Mpc}/h$  &  $\sim 0.5$ s\\
    $\xi(r,\mu)$ &(den+rsd+len) &up to $\sim 1000 \, \text{Mpc}/h$  & $\sim 17$ s\\
    \hline
      $\xi_\ell(r)$ &(den+rsd) &up to $\sim 1000 \, \text{Mpc}/h$  & $\sim 9$ s\\
        $\xi_\ell(r)$ &(den+rsd+len) &up to $\sim 1000 \, \text{Mpc}/h$  & $\sim 2$ min\\
        \hline
  $\Xi_\ell(r)$ &(den+rsd) & $\delta z =0.3$  & $\sim 1.5$ s\\
        $\Xi_\ell(r)$ &(den+rsd+len) &$\delta z =0.3$  & $\sim 3$ min\\
\hline
$\text{cov}^{(\xi)}_{\ell\ell'}(r_i,r_j)$ & ($N_p = 50$) & $N_p \times N_p$  & $\sim 20$ s\\ 
$\text{cov}^{(\Xi)}_{\ell\ell'}(r_i,r_j)$ &($N_p = 50$) &$N_p \times N_p$  & $\sim 20$ s\\
\hline
\end{tabular}
 \caption{Run time of {\sc coffe} calculating the correlation function at fixed $\mu$ (for $\sim 200$ separations), one multipole $\ell$ for $\xi_\ell$ and $\Xi_\ell$ or the covariance matrices on one Intel(R) Core(TM) i7-7700 CPU @ 3.60GHz core. {\sc coffe} is however parallelized using the $\texttt{openMP}$ standard.}
  \label{Ttime}
\end{table}

%%%%%%%%%%%%%%%%%%%%%%%%%%%%%%%%%%What's next?%%%%%%%%%%%%%%%%%%%%%%%%%%%%%%%%%%%%%%%%%%%%
\section{Conclusion and outlook}\label{next}
In this paper we have presented a code to calculate the relativistic full-sky correlation function and its covariance matrix. As we have shown previously~\cite{Tansella:2017rpi}, relativistic effects and wide-angle contributions are of the same order and it is therefore inconsistent to consider one but not the other. Presently, the code uses the linear matter power spectrum and therefore is most relevant on large scales. This will be important for the many planned future deep wide-angle surveys.  We have argued that the correlation function is a better tool than the --in principle equivalent-- angular power spectrum $C_\ell(z_1,z_1)$ for spectroscopic surveys.  As an example of how to use the code we have computed the signal-to-noise of the lensing term for some near-future galaxy surveys. The code is publicly available at \url{https://cosmology.unige.ch/content/coffe} or at \url{https://github.com/JCGoran/coffe}.

We finally discuss some features that we plan to implement in upcoming versions of {\sc coffe}:
\begin{itemize}
\item \emph{{\sc class} integration}: We will integrate {\sc class} on top of {\sc coffe} so that with only one parameter file it will be possible to generate the matter power spectrum necessary for the 2pF computation and to obtain the desired correlation function output. This integration will be particularly useful for forecasts. 
\item \emph{Non-linearities}: {\sc coffe} \source{v.1.0} is a fully linear code. On the one hand this is justified by the fact that wide-angles and relativistic projections effects are most relevant at large scales. On the other hand it will be important in future versions to include the effects of non-linearities on the 2pF.  Especially lensing, which is an integrated effect where non-linearities close to the observer contribute, is always affected by non-linearities. One can of course, already in the present version, use the halo-fit matter power spectrum for non-integrated effects to mimic these non-linearities. But this is not really consistent as long as the linearized continuity equation is used to infer velocities.  At small-scales the velocity dispersion is responsible for the \emph{fingers-of-god} effect which can be modeled as a convolution of the real-space 2pF with the probability distribution for velocity along the LOS~\cite{Fisher:1994ks,2011MNRAS.417.1913R,Bianchi:2014kba}. Also at intermediate scales, both the position and the shape of the BAO peak are affected by non-linearities~\cite{Eisenstein:2006nj,Crocce:2007dt,Smith:2007gi}. In this sense a promising formalism is the one developed in~\cite{McCullagh:2014jsa} as it is mostly based on quantities already computed in the code. 
\item \emph{Bias}: Another important feature that ought to be implemented in future versions of the code is the generalisation of the simple redshift-dependent bias $b(z)$ to different contributions of the bias expansion and scale-dependent bias, as they are known to have  important effects on the 2pF~\cite{2010PhRvD..82j3529D}.
\item \emph{Curvature}: A generalisation of the code functionality which seems trivial at first sight is to allow for non-zero curvature $\Omega_K \ne 0$; however, from a technical point of view it has some challenges. As we discussed in section~\ref{theo}, the way the code computes the correlation function is based on the fact that we can analytically re-sum eq.~(\ref{resum}) via the addition theorem for spherical Bessel functions. The $j_\ell$'s appear in the Fourier-Bessel transform $\De_\ell$ as they are the radial part of the eigenfunctions $\mathcal{Q}_k(\bx)$ of the flat-space laplacian (i.e. $\mathcal{Q}_k(\bx)=\text{Exp}( i \bk \cdot \bx)$). The addition theorem can then be derived from the identity
\be
e^{i \bk \cdot \br} = e^{i \bk \cdot \bx_2}e^{-i \bk \cdot \bx_1}
\ee
where $\br = \bx_2-\bx_1$. For a manifold with constant curvature, expressions for the eigenfunctions can be obtained (one finds expressions equivalent to the plane wave expansion of the flat case but with the spherical Bessel replaced by hyperspherical Bessel function~\cite{DiDio:2016ykq}) and they do satisfy 
\be
\mathcal{Q}_k(\bx_2-\bx_1)=\mathcal{Q}_k(\bx_2)\mathcal{Q}_k(-\bx_1)
\ee
from which the addition theorem can be derived. Therefore, including curvature in {\sc coffe} is in principle straight forward, but it will require additional theoretical and coding efforts.
\item \emph{EFT of DE}: An interesting application of the code will be to study the effect of dark energy and modified gravity on the galaxy 2pF beyond the cosmological constant behaviour. On the one hand the code heavily relies on the $\La$CDM equations and modifying it to account for a different model will require a substantial rewriting of some portions of it. On the other hand we can explore all the dark energy and modified gravity models that contain one additional scalar degree of freedom with the \emph{effective field theory of dark energy} (EFT of DE)~\cite{Creminelli:2008wc,Gubitosi:2012hu,Bloomfield:2012ff}. We can in fact describe a range of models only using an handful of couplings, making it a very useful approach to constrain deviations from GR. If we limit ourself to Horndeski theories and fix a background history close to $\La$CDM we can parametrise the changes to the Poisson and the anisotropy equations by introducing two scale- and time-dependent quantities $\mu$ and $\Sigma$:
\be
k^2 \Psi =- \mu(z,k) \frac{3\Omega_m  \HH^2}{2 a} \de_c \,,
\label{modpoiss}
\ee
\be
\Phi/\Psi =\Sigma(z,k) \,.
\label{modspli}
\ee
A simple suitable parametrisation for the two couplings and applications of this idea for the CMB and the galaxies angular power spectrum can be found in~\cite{Salvatelli:2016mgy,Zhao:2008bn}. 
\item \emph{Multi-tracer observables}: it has been shown that by combining different tracers of the density field (e.g. two populations of galaxies with different biases), one can reduce cosmic variance~\cite{McDonald:2008sh} and improve the detectability of relativistic effects in the power spectrum~\cite{Yoo:2012se} and in the $C_\ell$'s~\cite{Lorenz:2017iez}. In the 2pF, correlating two populations of galaxies has the particularity to generate a dipole and octupole contribution~\cite{Bonvin:2013ogt,Bonvin:2015kuc,Gaztanaga:2015jrs,Hall:2016bmm}, which for symmetry reasons are absent in the case of one tracer. This dipole can be used to test the equivalence principle in a model-independent way and constrain modifications of gravity with relativistic effects~\cite{Bonvin:2018ckp}. An extension of the code to multiple tracers is therefore planed in the future. This will require the computation of odd multipoles and their covariance matrices.
\end{itemize}

\acknowledgments{We thank Martin Kunz for useful discussions and Thomas Hahn for assistance with the \code{CUBA} library. This work is supported by the Swiss National Science Foundation.}

%%%%%%%%%%%%%%%%%%%%%%%%%%%%%%%%%%Appendix%%%%%%%%%%%%%%%%%%%%%%%%%%%%%%%%%%%%%%%%%%%%
\vspace{1.6cm}
%{\Large\bf Appendix}

\appendix

\section{Estimators and Covariances}\label{a:est}
In this appendix we give some more details on the estimators used and we derive their covariances. We especially derive our estimator $\hat \Xi_\ell$ which is new.

\subsection{Estimators}
Let us start with the estimator of the multipole of the correlation function averaged of the redshift bin $[\bar z-\de z,\bar z+\de z]$ given in eq.~\eqref{e:xihat},
\be\label{ea:xihat}
\hat \xi_\ell (r,\bar z) = \beta_\ell \sum_{ij} \De_i \De_j \PP_\ell(\mu_{ij}) \delta_K(r_{ij}-r)\,.
\ee
Here $\De_i\equiv \De(\bx_i,z_i)$ is  the number counts per pixel of size $L_p^3$ and $\mu_{ij}=r_\pa/r_{ij}$ where $\br_{ij} =\bx_i-\bx_j$ and $r_\pa=\chi(z_i)-\chi(z_j)$.
We want to normalise this estimator such that in the continuum limit, $N=V/L_p^3 \ra \infty$, its expectation value is the multipole of the correlation function.
Let us compute its expectation value,
\bea\label{ea:xihat-exp}
\langle\hat \xi_\ell (r,\bar z)\rangle &=& \beta_\ell \sum_{ij} \langle\De_i \De_j\rangle \PP_\ell(\mu_{ij}) \delta_K(r_{ij}-r)\,.
\eea
We now perform the continuum limit and set $\bx_i=\by +\br/2$, $\bx_j=\by-\br/2$ and $\mu_{ij}=\mu$. The volume of a ring of radius $r$ with direction cosine $\mu$ to the outward direction is $2\pi rdr\, rd\mu=2\pi r^2drd\mu$. In our discrete sample we have to replace the infinitesimal scale $dr$ by the pixel size $L_p$. Hence one of the sums can by replaced by $\int d^3y/L_p^3$ while the other becomes  $2\pi r^2\int d\mu/L_p^2$. Since at fixed mean redshift $\bar z$ the correlation function does not depend on $\by$, the $y$-integration just contributes a volume factor.
Putting it all together we obtain
\be
\label{ea:xihat-exp-cint1}
\langle\hat \xi_\ell (r,\bar z)\rangle = \beta_\ell \frac{2\pi r^2 V}{L_p^5}\int_{-1}^1d\mu \xi(r,\mu,\bar z)\PP_\ell(\mu)\,.
\ee
Inserting the expansion
\be
\xi(r,\mu,\bar z) = \sum_{\ell=0}^\infty \xi_\ell(r,\bar z)\PP_\ell(\mu)
\ee
and making use of the orthogonality relation
$$
\int_{-1}^1d\mu \PP_\ell(\mu))\PP_{\ell'}(\mu) =\frac{2}{2\ell+1}\de_{\ell,\ell'}\,,
$$
we find
\be
\label{ea:xihat-exp-cint1}
\langle\hat \xi_\ell (r,\bar z)\rangle = \beta_\ell \frac{4\pi r^2 V}{L_p^5(2\ell+1)}\ \xi_\ell(r,\bar z)\,.
\ee
In order for this to estimate $\xi_\ell$ we must choose
\be\label{e:betal}
 \beta_\ell =  \frac{2\ell+1}{4\pi}\frac{L_p^5}{ r^2 V} \,.
\ee

Let us now turn to the more sophisticated estimator $\hat\Xi_\ell$ which does not just assign a global mean redshift inside our redshift bin but instead assigns to each pair its correct mean redshift. Since the redshift resolution of a spectroscopic survey can be very high, $10^{-3}$ or better, it may well be that there are only a few galaxy pairs with a fixed distance and the precise mean redshift in a bin (of width $2\de z\sim 0.1$ or so). Therefore we shall integrate over the bin with a given redshift distribution $W(z)$. If one wants to select a fixed redshift one can simply choose $W$ to be a delta function.

The quantities $\De_i$ are again the number counts per pixel of size $L_p^3$. For the continuum limit of eq.~\eqref{e:Xi} we therefore have to replace $$\sum_{ij}\De_i\De_j \longrightarrow \int \frac{d^3x_id^3x_j}{L_p^6}\De(\bx_i)\De(\bx_j)\,,$$ where now $\De(\bx)$ is the continuum density contrast. We also replace the sum over discrete redshifts $z_k$ by an integral, $\int dz$. With this, the continuum limit of $\hat\Xi_\ell$ is
\be\label{ea:Xi}
\begin{split}
\hat \Xi_\ell (r,\bar z ,\de z) \ra \gamma_\ell &\int_{\bar z-\de z}^{\bar z+\de z}\hspace{-3mm} dz \,W(z) \int \frac{d^3x_i d^3x_j}{L_P^6} \bigg[ \frac{1+\cos\theta_{ij}}{2r_j^2} \\& \times \De(x_i,z_i) \De(x_j,z_j)\, \mathcal{P}_\ell(\mu_{ij})\, L_p\de(r_{ij}-r)\de(z_{ij}-z) \bigg] \,,
\end{split}
\ee
Note that we have replaced the Kronecker delta for the relative distance $r_{ij}$ by a Dirac delta multiplied by the pixel size. This not only has the right dimension but also takes care of the fact that we do not distinguish distances within one pixel.

We now make the coordinate transformation 
$$
\bx_i \ra \br=\bx_i-\bx_j \quad \mbox{ and } \quad \bx_j\ra \left(\chi_{ij}=\frac{|x_i|+|x_i|}{2}, \theta_j, \phi_j\right)\,.
$$
 Here $\theta_j$ and $\phi_j$ are the polar angles of $\bx_j$, and we do not distinguish between $\chi((z_i+z_j)/2)$ and $(|x_i|+|x_i|)/2$. The Jacobian of the transformation is readily calculated and amounts to
\be
J ={\rm Det}\left[\frac{\partial(\bx_i,\bx_j)}{\partial(\chi_{ij},\theta_j,\phi_j)}\right] = \frac{2x_j^2\sin\theta_j}{1+\cos\theta_{ij}}\,,
\ee
where as above $\theta_{ij}$ is the angle between $\bx_i$ and $\bx_j$. To eliminate this $x_j$ and $x_i$ dependent factor, we had to define our estimator as the density pair multiplied by the inverse of the factor. After this coordinate transformation the integration over $( \theta_j, \phi_j)$ can readily be performed and simply gives a factor $4\pi$. As above, the integral $d^3r$ can be written as 
$2\pi r^2drd\mu$.
We then obtain, for the expectation value of our estimator,
\be\label{ea:Xi-exp}
\langle\hat \Xi_\ell (r,\bar z ,\de z)\rangle =  \gamma_\ell\frac{(4\pi)^2r^2}{2L_p^5} \int_{\bar z-\de z}^{\bar z+\de z}\hspace{-3mm} dz \frac{W(z)}{H(z)} \int_{-1}^1d\mu \,\xi(r,\mu, z) \mathcal{P}_\ell(\mu)\,.
\ee
Here we have performed the $\chi_{ij}$-integration using the redshift delta-function and  the identity $\de(z_{ij}-z) = \de(\chi_{ij}-\chi)/H(z)$. Like above,  the $\mu$-integration  now yields the moment $\xi_\ell$,
\be\label{ea:Xi-ell--exp}
\langle\hat \Xi_\ell (x,\bar z ,\de z)\rangle =  \gamma_\ell\frac{(4\pi)^2r^2}{(2\ell+1)L_p^5} \int_{\bar z-\de z}^{\bar z+\de z}\hspace{-3mm} dz \frac{W(z)}{H(z)}  \xi_\ell(r, z)\,.
\ee
Hence in order to obtain the desired estimator for 
\be
\Xi_\ell(r,\bar z,\de z)= H_0\int_{\bar z-\de z}^{\bar z+\de z}\hspace{-3mm} dz \frac{W(z)}{H(z)}  \xi_\ell(r, z)\,,
\ee
we have to choose
\be
\ga_\ell = \frac{2\ell+1}{(4\pi)^2}\frac{L_p^5H_0 }{r^2 f_{\rm sky}} \,.
\ee
The factor $f_{\rm sky}$ has been introduced here to account also for partial sky coverage.
Note that the normalization factor $\ga_\ell$ has the correct dimension $($length$)^2$ to compensate for the dimensions of the factor $1/r_j^2$ in the sum of \eqref{ea:Xi} which therefore yields a dimensionless estimator. For the formula to hold, we also have assumed that the redshift window function is normalized to unity.

The estimators discussed here are optimal for the unrealistic case of a nearly full and homogeneous sky coverage. If there are certain parts of the sky where observations are better, more complete and or more precise, this can be taken into account by multiplying with an inhomogeneous weighting function in order to enhance the weight of these regions. Furthermore, for a complicated fractional sky coverage a simple multiplicative factor  $f_{\rm sky}$ is also not optimal.  In this paper we do not discuss these subtleties which, however are part of every real observation.

\subsection{Covariance matrix}
Here we briefly derive the expressions for the covariance matrices, Eqs.~\eqref{covxi} and \eqref{covXi}. Most of it can be found in the literature, see e.g. Refs~\cite{Bonvin:2015kuc,Hall:2016bmm,Tansella:2018tbo}, but in order to be more self contained we repeat the basic steps here. 

For the covariance matrix we only include the dominant terms: density and redshift space distortion. Even though at very large distance the correlation function is dominated by lensing, the covariance matrix $C(\br,\br')$ includes contributions from distances much smaller than $r$ and $r'$ where the standard terms largely dominate. This means that the density and RSD are the main contribution to the covariance matrix also at large distances. We also neglect redshift evolution and wide angle effects in the covariance matrix such that our estimator for the correlation function is
\be
\hat\xi(\br) =\frac{1}{V}\int_V d^3x\hat\De(\bx)\hat\De(\bx+\br)\,.
\ee

Including Poisson noise the observed two-point correlation function is given by
\be
\langle \hat\De(\bx) \hat\De(\bx')\rangle = \xi(\bx-\bx') +\frac{1}{\bar n}\de(\bx-\bx')\,,
\ee
where $\bar n$ is the mean number density in the redshift bin under consideration.

Assuming Gaussianity the covariance matrix of $\hat\xi$ is then becomes
\bea
C(\br,\br') &=& \langle \hat\xi(\br) \hat\xi(\br')\rangle -  \langle \hat\xi(\br)\rangle\langle \hat\xi(\br')\rangle\nonumber\\
&=& \frac{1}{V^2}\int_{V\times V} d^3xd^3x'\left[\xi(\bx-\bx')\xi(\bx+\br-\bx'-\br') +\xi(\bx+\br -\bx')\xi(\bx-\bx'-\br')\right]
\nonumber\\
&& +\frac{2}{V\bar n}\left[\xi(\br-\br')  + \xi(\br+\br')\right] +\frac{1}{\bar n^2}  \left[\de(\br-\br')  + \de(\br+\br')\right] \,.
\eea
Here we have used that the correlation function (for one population of galaxies) is symmetric, $\xi(\br)=\xi(-\br)$. The first line is the cosmic variance term, the second line contains terms which mix cosmic variance and Poisson noise and the last term is a pure Poisson noise term. Note that - as we have already anticipated - even if both $\br$ and $\br'$ are very large, the covariance matrix contains the correlation function at very small arguments, maybe in $\xi(\br-\br')$ but surely in the pure cosmic variance term, and these terms dominate the covariance. This has disadvantages, namely the covariance becomes much larger than  $\xi(\br)$ and $\xi(\br')$ for large $r$ and $r'$ leading to a small signal with large noise, but it also means that it is a good approximation to neglect wide angle effects and lensing in the covariance matrix as these are subdominant for small separation. 

After a change of variables, $(\bx,\bx')\ra (\bx,\by)$ with $\by =\bx-\bx'$, the $\bx$-integration of the cosmic variance term becomes trivial. Inserting the Fourier representation 
$$\xi(\by)= (2\pi)^{-3}\int d^3k P(\bk)\exp[i\bk\cd\by]\,,$$
 the $y$-integration of the cosmic variance term can be performed leading to a Dirac delta of the two Fourier variables. Representing also the Dirac delta of the Poisson term in Fourier space, we end up with
\bea
C(\br,\br') &=& \frac{1}{V(2\pi)^3}\int d^3k\left[P^2(\bk)+\frac{2}{\bar n}P(\bk)+\frac{1}{\bar n^2}\right] \left(e^{i\bk\cd(\br-\br')} +e^{i\bk\cd(\br+\br')}\right)\,. \label{ea:cov}
\eea

We now use the fact that (in the flat sky approximation) $P(\bk)=P(k,\nu)$ where $\nu$ is the direction cosine between the observation direction $\bn$ and $\bk$.
Furthermore, we write the exponentials in terms of Bessel functions, $j_\ell$, and Legendre polynomials $\PP_\ell$ as
$$
\exp(i\bk\cd\br) = \sum_{\ell=0}^\infty i^{\ell}(2\ell+1)\PP_\ell(\mu)j_\ell(kr)\,,
$$
where $\mu$ is the direction cosine between $\bk$ and $\br$. With this 
for example the pure cosmic  variance term becomes
 \bea
 \frac{1}{V(2\pi)^3}\!\int\!\! d^3kP^2(k,\nu)\sum_{\ell,\ell'}\PP_\ell(\mu)\PP_{\ell'}(\mu')j_\ell(kr)j_{\ell'}(kr')\left[i^{\ell-\ell'}\! \!+i^{\ell+\ell'}\right]\!(2\ell+1)(2\ell'+1) \,.
\eea
Since $P(\bk)$ is even in $\bk$, we obtain  non-vanishing results only if both, $\ell$ and $\ell'$ are even. Therefore $i^{\ell-\ell'}+i^{\ell+\ell'} =2i^{\ell-\ell'}$.
We expand also $P$ and $P^2$ in Legendre polynomials using $P(k,\nu)= P(k)(c_0 + c_2\PP_2(\nu)+c_4\PP_4(\nu))$ and
\be
P^2(k,\nu)=P^2(k)\sum_{\ell=0}^4\tilde c_{2\ell}\,\PP_{2\ell}(\nu)\,,
\ee
where the coefficients $\tilde c_L$ are obtained by expanding the square $(P(k,\nu)/P(k))^2$ in Legendre polynomials,
\be
\left(c_0\PP_0 + c_2\PP_2 + c_4\PP_4\right)^2  =\sum_{L=0}^8 \tilde c_L\PP_L \,.
\ee
The values $c_L$ and $\tilde c_L$ are given in \eqref{e:c0} to  \eqref{e:c4} and  \eqref{e:c0tilde} to \eqref{e:c8}.

Employing the addition theorem of spherical harmonics for $\nu =\hat\bk\cd\bn$, $\mu=\hat\bk\cd\hat\br$ and $\mu'=\hat\bk\cd\hat\br'$  we convert the Legendre polynomials into products of spherical harmonics. The angular integral of the pure covariance and of the mixed term leads to an angular integral of a product of three  spherical harmonics which can be performed exactly using~\cite{AW}
\be
\int d\Om_\bk  Y_{LM}(\hat\bk) Y_{\ell'm'}(\hat\bk)Y_{\ell m}(\hat\bk) = \sqrt{\frac{(2L+1)(2\ell'+1)(2\ell+1)}{4\pi}}\tj{L}{\ell'}{\ell}{0}{0}{0}\tj{L}{\ell'}{\ell}{M}{m'}{m}\,.
\ee
This yields
\bea
\int d\Om_\bk\PP_L(\nu)\PP_\ell(\mu)\PP_{\ell'}(\mu') &=& \sqrt{\frac{(4\pi)^5}{(2L+1)(2\ell'+1)(2\ell+1)}}\tj{L}{\ell'}{\ell}{0}{0}{0} \times \nonumber \\
&&\hspace{-3cm}\sum_{M,m,m'}\tj{L}{\ell'}{\ell}{M}{m'}{m}Y^*_{LM}(\bn) Y^*_{\ell'm'}(\hat\br')Y^*_{\ell m}(\hat\br) \,.
\eea
We now choose $\bn=\mathbf{e}_z$ so that $Y_{LM}(\bn) = \sqrt{(2L+1)/4\pi}\de_{M,0}$. 
Inserted this in \eqref{ea:cov} we find
\bea
C(\br,\br') &=& \frac{2(4\pi)^2}{V(2\pi)^3}\sum_{L,\ell,\ell',m,m'}\sqrt{(2\ell'+1)(2\ell+1)}\tj{L}{\ell'}{\ell}{0}{0}{0}i^{\ell-\ell'}\tj{L}{\ell'}{\ell}{0}{m'}{m} \times \nonumber \\
&&\hspace{-1.5cm}Y^*_{\ell'm'}(\hat\br')Y^*_{\ell m}(\hat\br)\int\! dkk^2\left[\tilde c_LP^2(\bk)+c_L\frac{2}{\bar n}P(\bk)+\de_{0L}\frac{1}{\bar n^2}\right]j_\ell(kr)j_{\ell'}(kr')\,. \label{ea:cov2}
\eea
 The covariance matrix for the multipoles $n$ and $n'$ is given by
 \bea
\text{cov}^{(\xi)}_{n,n'}(r,r') &=& \frac{(2\ell+1)(2\ell'+1)}{4}\int_{-1}^1d\mu\int_{-1}^1d\mu'\PP_{n}(\mu)\PP_{n'}(\mu')C(\br,\br')\,,
 \eea
 where now $\mu=\bn\cd\hat\br=\cos\theta$ and $\mu'=\bn\cd\hat\br'=\cos\theta'$.
 Using that $\PP_n(\mu) =\sqrt{4\pi/(2n+1)}Y_{n0}(\hat\br)$ together with the orthonormality of the spherical harmonics we can write
  \bea
 \text{cov}^{(\xi)}_{\ell,\ell'}(r,r') &=& \frac{1}{V\pi^2}(2\ell'+1)(2\ell+1)\tj{L}{\ell'}{\ell}{0}{0}{0}^2i^{\ell-\ell'} \times \nonumber \\
&&\hspace{-1cm}\left(\int dkk^2\left[\tilde c_LP^2(\bk)+c_L\frac{2}{\bar n}P(\bk)+\de_{0L}\frac{1}{\bar n^2}\right]j_\ell(kr)j_{\ell'}(kr') \right)\,. \label{ea:covnn'}
\eea
 
Integrating the last term with
$$\int_0^\infty dkk^2kj_\ell(kr)j_{\ell'}(kr')  = \de_{\ell\ell'}\frac{\pi}{2r^2}\de(r-r')\,,$$ we finally obtain
\bea
  \text{cov}^{(\xi)}_{\ell,\ell'}(r,r') &=& \frac{i^{\ell-\ell'}}{V}\Bigg[\frac{(2\ell+1)^2}{2\pi\bar n^2 r^2}\de(r-r')\de_{\ell\ell'}+ \frac{1}{\bar n} \mathcal{G}_{\ell\ell'} (r,r',\bar z) \sum\limits_\sigma c_\sigma\tj{\ell}{\ell'}{\sigma}{0}{0}{0}^2\\
&&+ \mathcal{D}_{\ell\ell'} (r,r',\bar z) \sum\limits_\sigma \tilde c_\sigma\tj{\ell}{\ell'}{\sigma}{0}{0}{0}^2 \Bigg] \,, 
\eea
where $ \mathcal{G}_{\ell\ell'}$ and $ \mathcal{D}_{\ell\ell'}$ are given in Eqs. \eqref{e:Gll} and \eqref{e:Dll}. This is simply the continuum limit of
 eq.~\eqref{covxi}.

In order to find the 
 corresponding expressions for $\Xi$  we integrate the result obtained for $\xi$ over the redshift interval with the weight given in \eqref{e:Xicont}. This is of course not very precise since it does not take into account the exact mean redshift of the points $\bx,~\bx+\br$ and $\bx',~\bx'+\br'$, but these redshifts depend also on the directions of $\br$ and $\br'$ which would lead to very complicated expressions.
Furthermore, since the covariance matrix is dominated by small distances, we expect only very minor changes which we neglect.

\section{$X^n_\ell$ and $Z^n_\ell$ list}\label{appcoeff}

The full list of $X^n_\ell$ is given (where $b_1= b(z_1), f_2=f(z_2)$ etc.):
\begin{flalign*} 
&X_0^0 \big|_\text{den} = b_1 b_2\,, \\
&X_0^0 \big|_\text{rsd} =  f_1 f_2   \frac{1+2\cos^2\theta}{15 }\,, \\ 
&X_2^0 \big|_\text{rsd} = -\frac{f_1 f_2}{21}\left[1+11\cos^2\theta +\frac{18\cos\theta(\cos^2\theta-1)\chi_1\chi_2}{r^2}\right] \,,\\
&X_4^0 \big|_\text{rsd} = f_1 f_2 \left[\frac{4(3\cos^2\theta-1)(\chi_1^4+\chi_2^4)}{35r^4} +  \chi_1\chi_2 (3+\cos^2\theta)\frac{3 (3+\cos^2\theta)\chi_1\chi_2-8(\chi_1^2+\chi_2^2)\cos\theta }{35r^4} \right] \,,  \\ 
&X_0^2\big|_\text{d1} = \HH_1\HH_2 f_1 f_2 G_1 G_2  \frac{r^2\cos\theta}{3}\,,  \\ 
&X_2^2\big|_\text{d1} = - \HH_1\HH_2 f_1 f_2 G_1 G_2 \left((\chi_2-\chi_1\cos\theta)(\chi_1-\chi_2\cos\theta)+\frac{r^2\cos\theta}{3}\right) \,,   \\ 
&X_0^{4}\big|_\text{d2} = (3-f_{\text{evo}1})(3-f_{\text{evo}2})\, r^4\HH_1^2\HH_2^2 f_1 f_2 \,,  \\ 
&X_0^{4}\big|_\text{g1} = \frac{9\, r^4 \Omega_m^2}{4 a_1 a_2} (1+G_1)(1+G_2) \HH_0^4 \,,  \\ 
&X_0^{4}\big|_\text{g2} = \frac{9\, r^4 \Omega_m^2}{4 a_1 a_2} (5 s_1-2)(5s_2 -2) \HH_0^4 \,,  \\ 
&X_0^{4}\big|_\text{g3} = \frac{9\, r^4 \Omega_m^2}{4 a_1 a_2} (f_1-1)(f_2-1) \HH_0^4  \\ 
&X_0^0 \big|_\text{den-rsd} = \frac{b_1 f_2 }{3} \,,\\
&X_2^0 \big|_\text{den-rsd} = -b_1 f_2  \left(\frac{2}{3} -(1-\cos^2\theta)\frac{\chi_1^2}{r^2}\right) \,, \\ 
&X_1^1 \big|_\text{den-d1} = -b_1 f_2 \HH_2 G_2 (\chi_1 \cos \theta -\chi_2) \,, \\ 
&X_0^2 \big|_\text{den-d2} = (3-f_{\text{evo}2})\,r^2 b_1 f_2 \HH_2^2 \,, \\ 
&X_0^2 \big|_\text{den-g1} = -b_1\frac{3 \Omega_m}{2 a_2} (1+G_2) r^2\HH_0^2\,, \\ 
&X_0^2 \big|_\text{den-g2} = -b_1\frac{3 \Omega_m}{2 a_2} (5s_2-2) r^2\HH_0^2\,, \\ 
&X_0^2 \big|_\text{den-g3} = -b_1\frac{3 \Omega_m}{2 a_2} (f_2-1)  r^2\HH_0^2  \,,\\
&X_1^1 \big|_\text{rsd-d1} = f_1f_2 \HH_2 G_2  \frac{(1+2\cos^2\theta)\chi_2-3 \chi_1\cos\theta}{5} \,, \\
&X_3^1 \big|_\text{rsd-d1} = f_1f_2 \HH_2 G_2 \frac{(1-3\cos\theta)\chi_2^3 +\cos\theta(5+\cos^2\theta)\chi_2^2\chi_1-2(2+\cos\theta^2)\chi_2\chi_1^2 +2\chi_1^3\cos\theta}{5r^2} \,, \\ 
&X_0^2 \big|_\text{rsd-d2} = \frac{3-f_{\text{evo}2}}{3} f_1 f_2  r^2 \HH_2^2\,,\\
&X_2^2 \big|_\text{rsd-d2} = -(3-f_{\text{evo}2})  f_1f_2  \HH_2^2 \left(\frac{2}{3}r^2 -(1-\cos^2\theta)\chi_2^2\right)\,, \\ 
&X_0^2 \big|_\text{rsd-g1} = - \frac{ \Omega_m}{2 a_2} f_1  (1+G_2) r^2\HH_0^2\,, \\
&X_2^2 \big|_\text{rsd-g1} = \frac{3 \Omega_m}{2 a_2}  f_1 (1+G_2) \HH_0^2 \left(\frac{2}{3}r^2 -(1-\cos^2\theta)\chi_2^2\right)\,, \\ 
&X_0^2 \big|_\text{rsd-g2} = -\frac{ \Omega_m}{2 a_2}f_1(5s_2-2) r^2\HH_0^2 \,, \\
&X_2^2 \big|_\text{rsd-g2} =  \frac{3 \Omega_m}{2 a_2}f_1(5s_2-2) \HH_0^2 \left(\frac{2}{3}r^2 -(1-\cos^2\theta)\chi_2^2\right)\,, \\ 
&X_0^2 \big|_\text{rsd-g3} =  - \frac{ \Omega_m}{2 a_2} f_1(f_2-1) r^2\HH_0^2 \,, \\
&X_2^2 \big|_\text{rsd-g3} = \frac{3 \Omega_m}{2 a_2} f_1(f_2-1) \HH_0^2 \left(\frac{2}{3}r^2 -(1-\cos^2\theta)\chi_2^2\right)\,, \\ 
&X_1^{3} \big|_\text{d1-d2} =   -(3-f_{\text{evo}2}) \HH_1 \HH_2^2 f_1 f_2  \, r^2(\chi_2 \cos \theta -\chi_1)\,, \\ 
&X_1^{3} \big|_\text{d1-g1} =    \frac{3 \Omega_m}{2 a_2} \HH_0^2 \HH_1 f_1 (1+G_2) \, r^2(\chi_2 \cos \theta -\chi_1)\,, \\ 
&X_1^{3} \big|_\text{d1-g2} =    \frac{3 \Omega_m}{2 a_2} \HH_0^2 \HH_1 f_1 (5s_2-2)  \, r^2(\chi_2 \cos \theta -\chi_1)\,, \\ 
&X_1^{3} \big|_\text{d1-g3} =    \frac{3 \Omega_m}{2 a_2} \HH_0^2  \HH_1 f_1 (f_2-1) \, r^2(\chi_2 \cos \theta -\chi_1)\,, \\ 
&X_0^{4} \big|_\text{d2-g1} = -\frac{3(3-f_{\text{evo}1}) \,r^4\Omega_m}{2 a_2} \HH_0^2 \HH_1^2 f_1 (1+G_2)\,, \\ 
&X_0^{4} \big|_\text{d2-g2} =-\frac{3(3-f_{\text{evo}1}) \,r^4\Omega_m}{2 a_2} \HH_0^2 \HH_1^2 f_1 (5s_2-2)\,, \\ 
&X_0^{4} \big|_\text{d2-g3} = -\frac{3(3-f_{\text{evo}1}) \,r^4\Omega_m}{2 a_2} \HH_0^2 \HH_1^2 f_1 (f_2-1)\,, \\ 
&X_0^{4} \big|_\text{g1-g2} =\frac{9 \,r^4\Omega_m^2}{4 a_1 a_2} \HH_0^4 (1+G_1)(5 s_2-2)\,, \\ 
&X_0^{4} \big|_\text{g1-g3} = \frac{9 \,r^4\Omega_m^2}{4 a_1 a_2} \HH_0^4 (1+G_1) (f_2-1)\,, \\ 
&X_0^{4} \big|_\text{g2-g3} = \frac{9 \,r^4\Omega_m^2}{4 a_1 a_2} \HH_0^4 (5s_1-2)(f_2-1)\,. \\  
\end{flalign*}
where
\be
G(z)= \frac{\dot \HH}{\HH^2}+\frac{2-5s}{\chi \HH}+5 s - f_\text{evo} \,.
\ee
The full list of $Z^n_\ell$ is given (note that inside the integral: $r=r(\la,\la')$ as defined in Eqs.~(\ref{er1}),(\ref{er2})):
\begin{flalign*} 
&Z \big|_\text{len} = \frac{9 \Omega_m^2}{4}\HH_0^4\frac{(2-5s_1)(2-5s_2)}{\chi_1\chi_2} \int\limits_0^{\chi_1} \!d\la \int\limits_0^{\chi_2} \!d\la' \frac{(\chi_1-\la)(\chi_2-\la')}{\la \la'} \frac{D_1(\la)D_1(\la')}{a(\la)a(\la')} \bigg\{ \frac{2}{5} (\cos^2\theta-1) \la^2 \la'^2 I^0_0(r) \\
&\quad +\frac{4 r^2 \cos\theta \la \la'}{3} I^2_0(r) +\frac{4 \cos\theta \la \la' (r^2 +6 \cos\theta \la \la')}{15} I^1_1(r) +\frac{2(\cos^2\theta -1)\la^2\la'^2(2r^4 +3 \cos\theta r^2 \la \la')}{7 r^4} I^0_2(r) \\
& \quad +\frac{2 \cos\theta \la \la' \left(2 r^4 +12\cos\theta r^2 \la\la' +15 (\cos^2\theta-1)\la^2\la'^2 \right)}{15 r^2} I^1_3(r)\\
&\quad +\frac{(\cos^2\theta-1)\la^2\la'^2 \left(6 r^4 +30\cos\theta r^2\la\la' +35 (\cos^2\theta -1)\la^2\la'^2 \right)}{35r^4} I^0_4(r)\bigg\} \,,\\
&Z \big|_\text{g4} = 9 \Omega_m^2\HH_0^4 \frac{(2-5s_1)(2-5s_2)}{\chi_1\chi_2} \int\limits_0^{\chi_1} d\la \int\limits_0^{\chi_2} d\la' \frac{D_1(\la)D_1(\la')}{a(\la)a(\la')}r^4 I^4_0(r) \,,\\
&Z \big|_\text{g5} = 9 \Omega_m^2\HH_0^4 G_1 G_2 \int\limits_0^{\chi_1} d\la \int\limits_0^{\chi_2} d\la' \frac{D_1(\la)D_1(\la')}{a(\la)a(\la')} \HH(\la)\HH(\la')(f(\la)-1)(f(\la')-1)r^4 I^4_0(r) \,,\\
&Z \big|_\text{den-len} = - \frac{3\Omega_m}{2} b_1 \HH_0^2 \frac{2-5s_2}{\chi_2}D_1(z_1) \int\limits_0^{\chi_2} d\la \frac{\chi_2-\la}{\la}\frac{D_1(\la)}{ a(\la)} \bigg\{ 2\chi_1\la\cos\theta I^1_1(r) -\frac{\chi_1^2\la^2(1-\cos^2\theta)}{r^2} I^0_2(r)\bigg\} \,,\\
&Z \big|_\text{rsd-len} = \frac{3\Omega_m}{2} f_1 \HH_0^2 \frac{2-5s_2}{\chi_2}D_1(z_1) \int\limits_0^{\chi_2} d\la \frac{\chi_2-\la}{\la}\frac{D_1(\la)}{ a(\la)}  \bigg\{ \frac{ \la}{15} (\la-6 \chi_1 \cos\theta+3 \la \cos2 \theta)I^0_0(r)\\ \quad &-\la\frac{6 \chi_1^3 \cos\theta-\chi_1^2 \la \left(9 \cos ^2\theta+11\right)+\chi_1 \la^2 \cos\theta (3 \cos2 \theta+19)-2 \la^3 (3 \cos2 \theta+1)}{21 r^2}I^0_2(r)\\ \quad&-\frac{\la}{35r^4} \bigg[-4 \chi_1^5 \cos\theta-\chi_1^3 \la^2 \cos\theta (\cos2 \theta+7)+\chi_1^2 \la^3 \left(\cos ^4\theta+12 \cos ^2\theta-21\right)\\ \quad&-3 \chi_1 \la^4 \cos\theta (\cos2 \theta-5)-\la^5 (3 \cos2 \theta+1)+12 \chi_1^4 \la\bigg]I^0_4(r) \bigg\} \,,\\
&Z \big|_\text{d1-len} = \frac{3\Omega_m}{2} \HH_0^2 \HH_1 f_1 G_1\frac{2-5s_2}{\chi_2}D_1(z_1) \int\limits_0^{\chi_2} d\la \frac{\chi_2-\la}{\la}\frac{D_1(\la)}{ a(\la)}  \bigg\{\frac{2 \la}{15}\bigg(\cos\theta \left(\la^2-2 \chi_1^2\right) \\&\quad+\chi_1 \la (2 \cos2 \theta -1)\bigg) I^1_1(r) +\frac{2}{3}r^2 \la \cos\theta I^2_0(r)  \\&\quad-\la\frac{4 \chi_1^4 \cos\theta-\chi_1^3 \la \left(\cos ^2\theta+9\right)+\chi_1^2 \la^2 \cos\theta \left(\cos ^2\theta+5\right)-2 \chi_1 \la^3 (\cos2 \theta -2)-2 \la^4 \cos\theta}{15 r^2}I^1_3(r)  \bigg\} \,,\\
&Z \big|_\text{d2-len} = - \frac{3\Omega_m}{2}(3-f_{\text{evo}1})f_1\HH_1^2 \HH_0^2 \frac{2-5s_2}{\chi_2}D_1(z_1) \int\limits_0^{\chi_2} d\la \frac{\chi_2-\la}{\la }\frac{D_1(\la)}{ a(\la)}  \bigg\{ 2\chi_1\la r^2\cos\theta I^3_1(r) \\&\quad-\chi_1^2\la^2(1-\cos^2\theta)I^2_2(r) \bigg\} \,,\\
&Z \big|_\text{g1-len} = \frac{9\Omega_m^2}{4}(1+G_1) \HH_0^4 \frac{2-5s_2}{\chi_2}D_1(z_1) \int\limits_0^{\chi_2} d\la \frac{\chi_2-\la}{\la }\frac{D_1(\la)}{ a(\la)}  \bigg\{ 2\chi_1\la r^2\cos\theta I^3_1(r) \\&\quad-\chi_1^2\la^2(1-\cos^2\theta)I^2_2(r) \bigg\} \,,\\
&Z \big|_\text{g2-len} = \frac{9\Omega_m^2}{4}(5s_1-2) \HH_0^4 \frac{2-5s_2}{\chi_2}D_1(z_1) \int\limits_0^{\chi_2} d\la \frac{\chi_2-\la}{\la }\frac{D_1(\la)}{ a(\la)}  \bigg\{ 2\chi_1\la r^2\cos\theta I^3_1(r) \\&\quad-\chi_1^2\la^2(1-\cos^2\theta)I^2_2(r) \bigg\} \,,\\
&Z \big|_\text{g3-len} = \frac{9\Omega_m^2}{4}(f_1-1) \HH_0^4 \frac{2-5s_2}{\chi_2}D_1(z_1) \int\limits_0^{\chi_2} d\la \frac{\chi_2-\la}{\la }\frac{D_1(\la)}{ a(\la)}  \bigg\{ 2\chi_1\la r^2\cos\theta I^3_1(r) \\&\quad-\chi_1^2\la^2(1-\cos^2\theta)I^2_2(r) \bigg\} \,,\\
&Z \big|_\text{g4-len} = \frac{9\Omega_m^2}{2}\HH_0^4 \frac{(2-5s_1)(2-5s_2)}{\chi_1\chi_2}\int\limits_0^{\chi_1} d\la \int\limits_0^{\chi_2} d\la' \frac{\chi_2-\la'}{\la' }\frac{D_1(\la)D_1(\la')}{ a(\la)a(\la')}  \bigg\{ 2\la\la' r^2\cos\theta I^3_1(r) \\&\quad-\la^2\la'^2(1-\cos^2\theta)I^2_2(r) \bigg\} \,,\\
&Z \big|_\text{g5-len} = \frac{9\Omega_m^2}{2}\HH_0^4 G_1 \frac{2-5s_2}{\chi_2}\int\limits_0^{\chi_1} d\la \int\limits_0^{\chi_2} d\la' \HH(\la)(f(\la)-1)\frac{\chi_2-\la'}{\la' }\frac{D_1(\la)D_1(\la')}{ a(\la)a(\la')}  \bigg\{ 2\la\la' r^2\cos\theta I^3_1(r) \\&\quad-\la^2\la'^2(1-\cos^2\theta)I^2_2(r) \bigg\} \,,\\
&Z \big|_\text{den-g4} =-3\Omega_m\HH_0^2 b_1 \frac{2-5s_2}{\chi_2} D_1(z_1)\int\limits_0^{\chi_2} d\la\frac{D_1(\la)}{ a(\la)} r^2 I^2_0(r) \,,\\
&Z \big|_\text{den-g5} = -3\Omega_m\HH_0^2 b_1 G_2 D_1(z_1)\int\limits_0^{\chi_2} d\la  \HH(\la)(f(\la)-1)\frac{D_1(\la)}{ a(\la)} r^2 I^2_0(r)\,,\\
&Z \big|_\text{rsd-g4} =3\Omega_m\HH_0^2 f_1 \frac{2-5s_2}{\chi_2} D_1(z_1)\int\limits_0^{\chi_2} d\la\frac{D_1(\la)}{ a(\la)} \bigg\{\left(\frac{2 r^2}{3}+(\cos^2\theta-1)\la^2 \right) I^2_2(r)-\frac{r^2}{3} I^2_0(r) \bigg\} \,,\\
&Z \big|_\text{rsd-g5} =3\Omega_m\HH_0^2 f_1 G_2 D_1(z_1)\int\limits_0^{\chi_2} d\la \HH(\la)(f(\la)-1)\frac{D_1(\la)}{ a(\la)} \bigg\{\left(\frac{2 r^2}{3}+(\cos^2\theta-1)\la^2 \right) I^2_2(r)-\frac{r^2}{3} I^2_0(r) \bigg\} \,,\\
&Z \big|_\text{d1-g4} =3\Omega_m\HH_0^2 \HH_1f_1 \frac{2-5s_2}{\chi_2} D_1(z_1)\int\limits_0^{\chi_2} d\la\frac{D_1(\la)}{ a(\la)} \bigg\{ r^2(\la \cos\theta -\chi_1)I^3_1(r)\bigg\} \,,\\
&Z \big|_\text{d1-g5} =3\Omega_m\HH_0^2 \HH_1f_1G_2 D_1(z_1)\int\limits_0^{\chi_2} d\la  \HH(\la)(f(\la)-1) \frac{D_1(\la)}{ a(\la)} \bigg\{ r^2(\la \cos\theta -\chi_1)I^3_1(r)\bigg\} \,,\\
&Z \big|_\text{d2-g4} =-3\Omega_m\HH_0^2 (3-f_{\text{evo}1})f_1\HH_1^2 \frac{2-5s_2}{\chi_2} D_1(z_1)\int\limits_0^{\chi_2} d\la\frac{D_1(\la)}{ a(\la)} r^4 I^4_0(r) \,,\\
&Z \big|_\text{d2-g5} =-3\Omega_m\HH_0^2 (3-f_{\text{evo}1})f_1\HH_1^2 G_2 D_1(z_1)\int\limits_0^{\chi_2} d\la  \HH(\la)(f(\la)-1)  \frac{D_1(\la)}{ a(\la)} r^4 I^4_0(r) \,,\\
&Z \big|_\text{g1-g4} =\frac{9 \Omega_m^2}{2 a_1}\HH_0^4 (1+G_1)\frac{2-5s_2}{\chi_2} D_1(z_1)\int\limits_0^{\chi_2} d\la\frac{D_1(\la)}{ a(\la)} r^4 I^4_0(r) \,,\\
&Z \big|_\text{g1-g5} =\frac{9 \Omega_m^2}{2 a_1}\HH_0^4 (1+G_1)G_2 D_1(z_1)\int\limits_0^{\chi_2} d\la \HH(\la)(f(\la)-1)  \frac{D_1(\la)}{ a(\la)} r^4 I^4_0(r) \,,\\
&Z \big|_\text{g2-g4} =\frac{9 \Omega_m^2}{2 a_1}\HH_0^4 (5s_1-2)\frac{2-5s_2}{\chi_2} D_1(z_1)\int\limits_0^{\chi_2} d\la\frac{D_1(\la)}{ a(\la)} r^4 I^4_0(r) \,,\\
&Z \big|_\text{g2-g5} =\frac{9 \Omega_m^2}{2 a_1}\HH_0^4 (5s_1-2)G_2 D_1(z_1)\int\limits_0^{\chi_2} d\la \HH(\la)(f(\la)-1)  \frac{D_1(\la)}{ a(\la)} r^4 I^4_0(r) \,,\\
&Z \big|_\text{g3-g4} =\frac{9 \Omega_m^2}{2 a_1}\HH_0^4 (f_1-1)\frac{2-5s_2}{\chi_2} D_1(z_1)\int\limits_0^{\chi_2} d\la\frac{D_1(\la)}{ a(\la)} r^4 I^4_0(r) \,,\\
&Z \big|_\text{g3-g5} =\frac{9 \Omega_m^2}{2 a_1}\HH_0^4 (f_1-1)G_2 D_1(z_1)\int\limits_0^{\chi_2} d\la \HH(\la)(f(\la)-1)  \frac{D_1(\la)}{ a(\la)} r^4 I^4_0(r) \,,\\
&Z \big|_\text{g4-g5} =9 \Omega_m^2 \HH_0^4 G_2 \frac{2-5s_1}{\chi_1} \int\limits_0^{\chi_1} d\la \int\limits_0^{\chi_2} d\la' \HH(\la')(f(\la')-1)\frac{D_1(\la)D_1(\la')}{ a(\la)a(\la')} r^4 I^4_0(r) \,.\\
\end{flalign*} 

%\newpage
\bibliographystyle{utcaps}
\bibliography{coffe-refs}

\end{document}